\documentclass[aip,superscriptaddress,amsmath,amssymb,reprint]{revtex4-1}

\usepackage{graphicx}
\graphicspath{{pictures/}}
\usepackage{dcolumn}
\usepackage{bm}

\usepackage[utf8]{inputenc}
\usepackage[T1]{fontenc}
\usepackage{mathptmx}
\usepackage{etoolbox}

\usepackage{xcolor}
\usepackage{booktabs}
\usepackage{siunitx}
\usepackage{braket}
\usepackage{hyperref}
\usepackage{cleveref}
\usepackage{bbold}
\usepackage{ulem}
\usepackage{multirow}

\definecolor{lightblue}{RGB}{68,114,196}
\definecolor{darkgreen}{RGB}{0,100,0}

\DeclareSIUnit\angstrom{\text {Å}}

\begin{document}

\preprint{AIP/123-QED}
\sffamily
\title[Nonadiabatic EET through dendrimer building blocks]{Excitation-Energy Transfer and Vibronic Relaxation through Light-Harvesting Dendrimer Building Blocks: a Nonadiabatic Perspective}

\author{Joachim Galiana}
\author{Benjamin Lasorne}
\email{benjamin.lasorne@umontpellier.fr}
\affiliation{\textsuperscript{\textnormal{1)}}ICGM, Univ Montpellier, CNRS, ENSCM, Montpellier, France}

\date{\today}
\normalem
\begin{abstract}

\noindent
The light-harvesting excitonic properties of poly(phenylene ethynylene) (PPE) extended dendrimers (tree-like $\pi$-conjugated macromolecules) involve a directional cascade of local excitation-energy transfer (EET) processes occurring from the ``leaves'' (shortest branches) to the ``trunk'' (longest branch), which can be viewed from a vibronic perspective as a sequence of internal conversions occurring among a connected graph of nonadiabatically coupled locally-excited (LE) electronic states via conical intersections.
The smallest PPE building block able to exhibit EET, the asymmetrically \emph{meta}-substituted PPE oligomer with one acetylenic bond on one side and two parallel ones on the other side (hence, 2-ring and 3-ring \emph{para}-substituted pseudo-fragments), is a prototype and the focus of the present work.
From linear-response time-dependent density-functional theory (LR-TD-DFT) electronic-structure calculations of the molecule as regards its first two nonadiabatically coupled, optically active, singlet excited states, we built a (1+2)-state-8-dimensional vibronic-coupling Hamiltonian (VCH) model for running subsequent multiconfiguration time-dependent Hartree (MCTDH) wavepacket relaxations and propagations, yielding both steady-state absorption and emission spectra, as well as real-time dynamics. 
The EET process from the shortest branch to the longest one occurs quite efficiently (about 80\% quantum yield) within the first \qty{25}{\femto\second} after light excitation and is mediated vibrationally through acetylenic and quinoidal bond-stretching modes together with a particular role given to the central-ring anti-quinoidal rock-bending mode. Electronic and vibrational energy relaxations, together with redistributions of quantum populations and coherences, are interpreted herein through the lens of a nonadiabatic perspective, showing some interesting segregation among the foremost photoactive degrees of freedom as regards spectroscopy and reactivity.
\end{abstract}

\maketitle
\section{Introduction}
\label{sec:introduction}
The qualitative and quantitative modeling of excitation-energy transfer (EET) processes occurring through macro/supra-molecules or materials is of much interest from both chemical and physical viewpoints, with potential applications as regards light-harvesting properties ranging for photovoltaic technologies to artificial photosynthesis.\cite{balzani_harvesting_1995,gilat_light_1999,adronov_light-harvesting_2000,balzani_light-harvesting_2003}

Time-dependent density functional theory (TD-DFT) is now considered as an efficient and often reliable computational method for probing the excited-state manifolds of large organic molecules (about 10 to 100 atoms) and has been assessed against systematic benchmarks\cite{adamo_calculations_2013,knysh_excess_2023}. This approach has made it possible for theoretical photochemistry to address systems of realistic interest for technological applications.

Theoretical and computational studies of EET have historically relied on perturbative approximations involving tight-binding, site-based, and electric-dipole-type light-matter interaction approaches, so as to address in particular Frenkel-type excitonic Hamiltonian models (see, for example, Förster/Dexter approaches and Redfield equations\cite{yang_influence_2002,yang_reduced_2005}).

More recently, more sophisticated, non-perturbative, open-quantum approaches making use of the `Hierarchical Equations of Motion' (HEOM) technique have been developed to evaluate EET dynamics within various system-bath partitions of such nonadiabatic molecular systems.\cite{yan_theoretical_2021,breuil_funneling_2021}.

It must be noted, however, that the detailed real-time dynamics of EET remains a computational challenge to date as regards fully atomistic simulations. From a vibronic and chemical perspective, EET naturally involves nuclear/vibrational dynamics through several high-dimensional potential energy surfaces (PESs) related to nonadiabatically coupled electronic excited states. 
EET may thus be modelled from a chemical viewpoint via the concept of ultrafast internal conversion, along a ``reaction'' path going through a sequence of conical intersections.\cite{domcke_conical_2004,domcke_conical_2011}

In this work, we focus on the study of EET within a prototypical light-harvesting antennae building block.
Our molecule of interest is an oligo(phenylene ethynylene) (PPE). It is one of the basic units of the macromolecular nano-star, first synthesized in 1994 by Moore and co-workers.\cite{xu_phenylacetylene_1994}
The latter is an example of an extended PPE dendrimer (tree-shaped macromolecule), built on PPE oligomers of increasing length from the periphery to the core.

In such systems, it has been shown that the ultrafast and highly efficient EET process is to be interpreted as the consequence of a unidirectional excitation-energy gradient from the shortest (high optical energy gap) to the longest (low optical energy gap) $\pi$-conjugated branches.\cite{devadoss_energy_1996,shortreed_directed_1997,kleiman_ultrafast_2001,melinger_optical_2002} 

Early attempts for modeling and understanding EET have relied so far on kinetic models such as Förster exchange and were essentially based on the Frenkel-exciton Hamiltonian, allowing seminal computations of the absorption spectrum of the nano-star.
They confirmed indeed that the features of the spectrum could be attributed to the additive involvement of locally-excited (LE) electronic states on the various branches\cite{kopelman_spectroscopic_1997,swallen_dendrimer_1999,swallen_correlated_2000,rana_steady-state_2001,wu_exciton_2006,palma_electronic_2010} with little or no charge transfer among them.

\begin{figure}
  \includegraphics[width=0.8\linewidth]{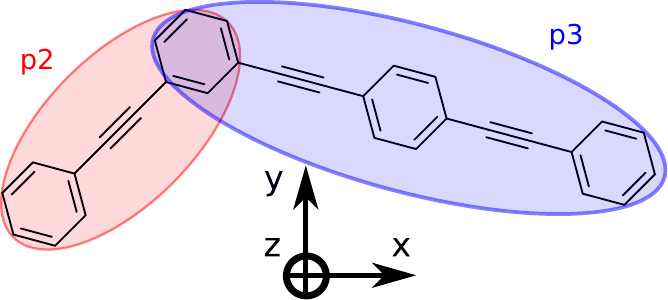}
  \caption{
  The asymmetrical \textit{meta}-substituted PPE, m23, decomposed as a dyad of linear 2-ring (p2) and 3-ring (p3) pseudo-fragments, sharing one central phenylene ring. Our choice of Cartesian axes is indicated for further reference.
  }
  \label{fig:m23_fragments}
\end{figure}

Fernandez-Alberti and co-workers then proposed the first time-resolved simulations of some initially excited building blocks of the nano-star, in particular for basic species made of \textit{meta}-substituted 2-, 3-, and 4-ring PPEs (the simplest combination of 2- and 3-ring species -- the focus of the present work -- will be called m23 hereafter; \cref{fig:m23_fragments}).

These authors discussed the key role played by the acetylenic bonds in PPE oligomers (among them, m23) for EET, based on semi-empirical and mixed quantum-classical trajectory-based molecular dynamics simulations.\cite{fernandez-alberti_nonadiabatic_2009,fernandez-alberti_unidirectional_2010,fernandez-alberti_non-adiabatic_2016,malone_nexmd_2020}
Their direct-dynamics simulations unveiled ultrafast EET (within \qty{30}{\femto\second} after excitation) involving fast access to a seam of conical intersection. 
Further modeling led to a first description of the nonadiabatic coupling vectors in terms of normal modes of vibration, with the identification of state-specific vibrations for EET.\cite{soler_analysis_2012,fernandez-alberti_identification_2012}
It is to be noted that the transient absorption spectra of m23 has also been evaluated recently with mixed quantum-classical trajectory-based molecular dynamics simulations, and interpreted as a spectral fingerprint of EET.\cite{hu_spectral_2021}

We hereby propose a complementary, and expectedly more quantitative, wavepacket quantum dynamics study of the first asymmetrical building-block m23, started from the Franck-Condon (FC) region (sudden approximation) and designed for exploring over time the region around the Minimum Energy Conical Intersection (MECI) between the first two singlet electronic excited states.
Our aim is to characterize the reaction pathway adequately with respect to the most active normal modes of vibration of the molecule, and this  based on quantum-chemistry TD-DFT calculations, for which numerical accuracy has already been assessed.\cite{ho_vibronic_2017,ho_diabatic_2019}

We propose a model able to describe the adiabatic potential-energy surfaces (PESs) of the first two nonadiabatically coupled singlet electronic excited states of the molecule, $\text{S}_{1}$ and $\text{S}_{2}$, \textit{via} a so-called Linear Vibronic Coupling (LVC) Hamiltonian (quasi-)diabatic model (diabatization by ansatz), which can be viewed as a version of the Frenkel Hamiltonian with explicit nuclear-displacement dependence.

Our parameterized LVC model has been used here for running wavepacket quantum dynamics with the now-standard Multiconfiguration Time-Dependent Hartree (MCTDH) approach.\cite{beck_multiconfiguration_2000}
We previously used the same strategy for simulating the early time-resolved evolution and for calculating the stationary absorption and emission UV-visible spectra of the first symmetrical \textit{meta}-substituted PPE with two diphenylacetylene (DPA) p2 branches, namely m22, within a three-dimensional LVC model.\cite{galiana_unusual_2023}

Because setting a vibronic-coupling Hamiltonian model is not evident for high-dimensional situations in the absence of molecular-symmetry mode segregation (based on point-group properties), an effort was made here as regards the choice of the explicitly included degrees of freedom and how they should enter the model Hamiltonian matrix entries (off-diagonal coupling or diagonal tuning; linear or quadratic coordinate terms).

\begin{figure}
\includegraphics[width=0.95\linewidth]{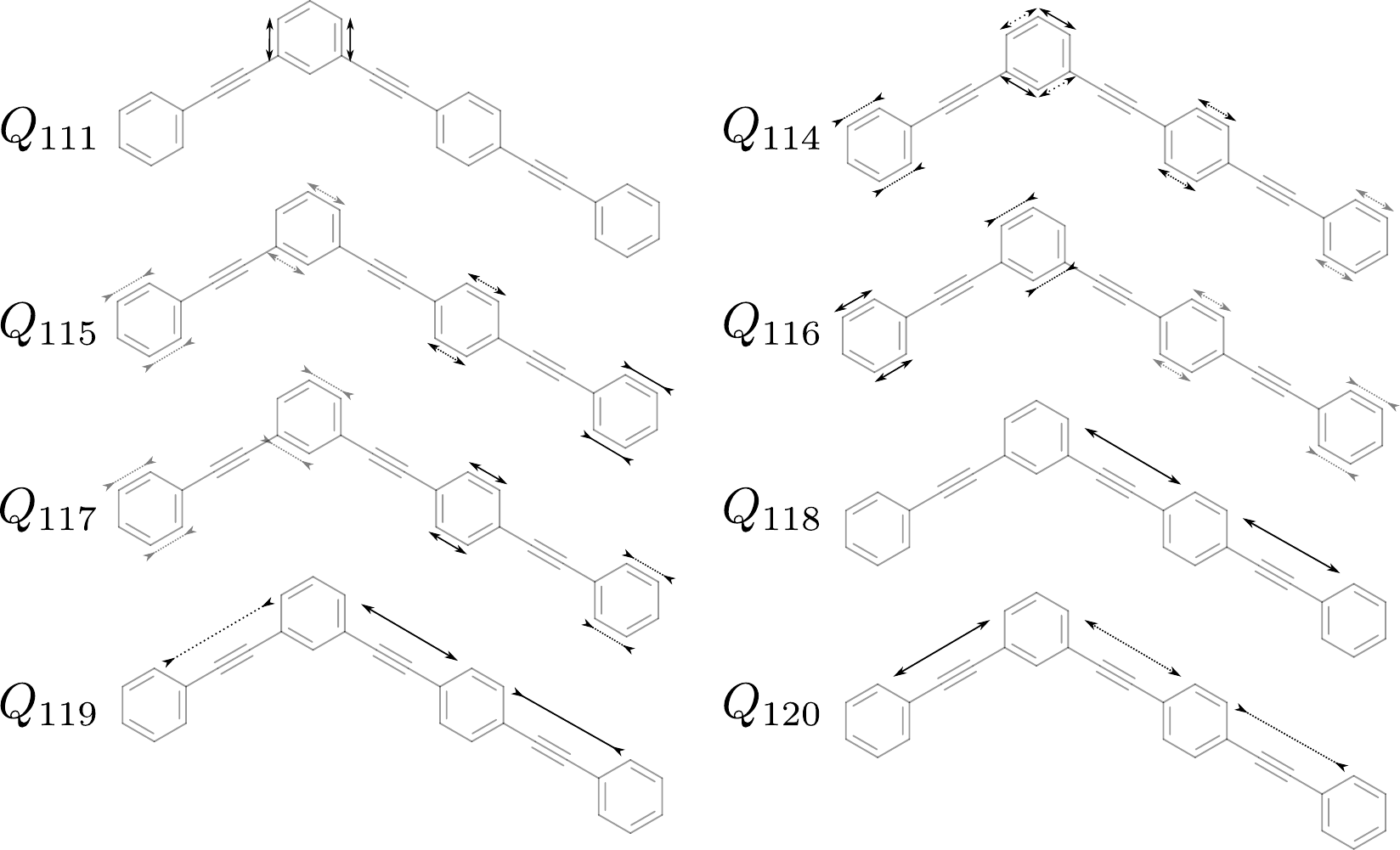}
\caption{Schematic representation of the nuclear displacements associated to the 8 selected normal modes of vibration from the equilibrium geometry of m23 in its electronic ground state, $\text{S}_{0}$. 
  Quantitative nuclear displacements in the real Cartesian space are given in the SI (fig. SI1).
  }
\label{fig:quinace_modes_schematic}
\end{figure}

Quite obviously, active modes should describe significant changes in molecular geometries occurring during EET between $\text{S}_{2}$ and $\text{S}_{1}$. As it stands, we chose a reduced 8-dimensional (8-D) and (1+2)-state model as a minimal description, with the most prominent 5 quinoidal and 3 acetylenic modes among a total of 138 (\cref{fig:quinace_modes_schematic}).

The 8 normal modes of vibration that we selected are partially localized on one or the other branches (the p2 and p3 pseudo-fragments). They can be transformed into 8 local modes on the phenylene sites (5 local modes) and acetylene sites (3 local modes).

The choice of the 3 acetylenic modes was motivated by the obvious and prominent role played by the acetylenic bonds in the EET process of m23.\cite{fernandez-alberti_nonadiabatic_2009}
The choice for the other 5 quinoidal modes was made so as to provide a better description of the initial gradients, together with the conical intersection seam and branching plane, along the line of what we observed for m22.\cite{galiana_unusual_2023} This will be illustrated with various considerations hereafter.

Our paper is organized as follows. In \cref{sec:computational_details}, we provide relevant computational details for quantum chemistry calculations and quantum dynamics simulations of m23, and we describe the LVC Hamiltonian model that was used for the nonadiabatically coupled PESs.
In \cref{sec:results_stationary} we describe the procedure for building the corresponding LVC Hamiltonian and its parametrization from \textit{ab initio} data, and discuss its validity.
In \cref{sec:results_dynamics}, we analyze the early dynamics simulations induced by two different types of excitation, with emphasis on the population kinetics and the stationary UV-visible spectra.
Then, we propose a mode-specific analysis of the EET phenomenon within our reduced model, and complete it with a discussion on internal coordinates to compare with the existing literature.
Finally, we initiate a discussion on the internal vibrational (or vibronic, here) redistribution of the excitation energy.
Conclusions and prospects for future works are made in \cref{sec:conclusion}

\section{Computational and modeling details}
\label{sec:computational_details}
\subsection{Quantum Chemistry and Electronic Structure}
\label{sec:quantum_chemistry}
Our molecular system of interest here is the left-right-asymmetrical but planar \textit{meta}-substituted PPE oligomer named m23, with point group symmetry C\textsubscript{s}.
Quantum chemistry calculations were performed using the Gaussian16 package (revision A.03)\cite{g16} using density-functional theory (DFT) for the electronic ground state and linear-response time-dependent DFT (TD-DFT) for the electronic excited states, at the CAM-B3LYP/6-31+G\textsuperscript{*} level of theory.

Other PPE  oligomers have already been studied at this level of theory, which was found to be suitable, for both \textit{para}- and \textit{meta}-substituted species.\cite{adamo_calculations_2013,ho_vibronic_2017,ho_diabatic_2019,galiana_unusual_2023}
The minima of the electronic ground state MinS\textsubscript{0} (also called Franck-Condon geometry) and of the first two singlet electronic excited states, MinS\textsubscript{1,2}, were optimized and characterized with frequency calculations. 
Relevant data are summarized in the first three rows of \cref{tab:critical_points}.
\begin{table}[h]
  \caption{
    Energies in \unit{\electronvolt} of the first two adiabatic excited states at the critical points in the \textit{ab initio} PESs (at the CAM-B3LYP/6-31+G\textsuperscript{*} level of theory) and in the LVC model PESs, described in \cref{sec:results_stationary}.
    The detailed positions of the points in the reduced-dimension model are given in the SI (Table SI2).}
  \label{tab:critical_points}
  \begin{tabular}{lcccc}
    \toprule
    Critical Point & MinS\textsubscript{0} & MinS\textsubscript{1} & MinS\textsubscript{2} & MECI(S\textsubscript{1}/S\textsubscript{2}) \\
    \midrule
    Optimization & & & & \\
    $E(\text{S}_{1})$ & 3.88 & 3.61 & 3.99 & 4.30 \\
    $E(\text{S}_{2})$ & 4.45 & 4.62 & 4.17 & 4.30 \\
    LVC Model & & & & \\
    $V^{(1)}$ & 3.88 & 3.67 & 4.03 & 4.41 \\
    $V^{(2)}$ & 4.45 & 4.59 & 4.24 & 4.41 \\
    \bottomrule
  \end{tabular}
\end{table}

Our reference set of normal modes of vibration throughout the paper will be those computed at MinS\textsubscript{0} (Franck-Condon geometry). 
The corresponding nuclear displacements are schematized in \cref{fig:quinace_modes_schematic} and more rigorously represented in SI (fig. SI1).

Note that, for comparison, a few similar calculations were made and exploited for the ground and first singlet electronic excited states of the isolated linear pseudofragment oligomers with two and three rings, named p2 and p3 hereafter.

\begin{figure*}
  \includegraphics[width=0.9\linewidth]{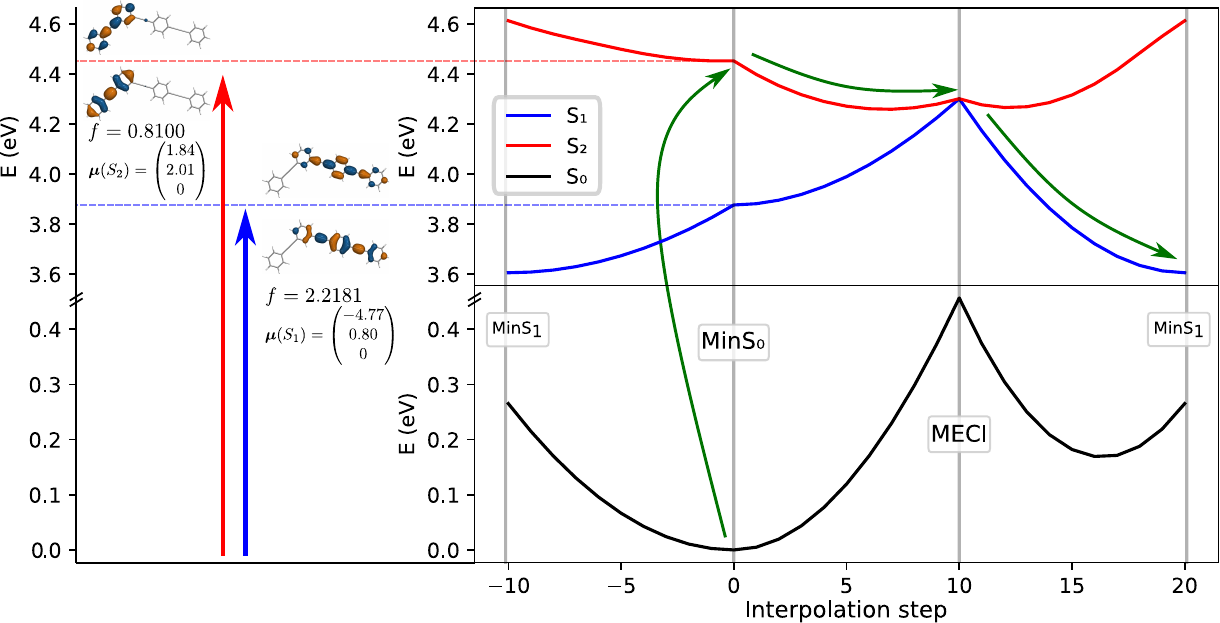}
  \caption{
    Left: oscillator strengths, electronic transition dipole moments, and natural-transition orbitals for the first two vertical transitions at the minimum of S\textsubscript{0}.
    Right: adiabatic energies of the electronic ground state and of the first two excited singlet states along linear interpolations between the minimum of S\textsubscript{1} (both at \num{-10} and \num{20}), the minimum of S\textsubscript{0} (at \num{0}), and the MECI (at \num{10}).}
  \label{fig:interpolation_etdm}
\end{figure*}

Let us provide some details now. At the $\text{S}_{0}$ Franck-Condon (FC) geometry, the first and second singlet electronic excited states, $\text{S}_{1}$ and $\text{S}_{2}$, are locally-excited (LE), single-electron-transitions of $\pi-\pi^*$ nature, localized on either the p3 or p2 pseudofragments, respectively (blue and red arrows in \cref{fig:interpolation_etdm}, left). They really are excitonic prototypes.

Both are optically bright (with significant oscillator strengths: $f(\text{S}_{1})=2.22$ and $f(\text{S}_{2})=0.81$) and of symmetry A'. Note that the oscillator strengths do not vary significantly from the FC geometry to the minima of both excited states: at the minimum of S\textsubscript{1}, $f(\text{S}_{1})=2.23$; at the minimum of S\textsubscript{2}, $f(\text{S}_{2})=0.53$. This shows that their ``diabatic'' character is preserved over both pathways.

More specifically, the first two vertical transitions at the FC geometry are dominated by HOMO/LUMO and HOMO-1/LUMO+1 transitions, localized -- as expected -- on p3 and p2, respectively (see \cref{fig:interpolation_etdm}). For each state, the most important pair of natural transition orbitals (weight $> 90\%$) involves a transition from a $\pi$ orbital to a $\pi^{*}$ on the same p3/p2-pseudofragment, consistent with an ``alternated-aromatic to cumulenic-quinoidal'' excited rebonding pattern with respect to the ground-state.

The Minimum Energy Conical Intersection (MECI) between the $\text{S}_{1}$ and $\text{S}_{2}$ states has been optimized using a modified implementation of the hybrid combined-gradient/composed-step algorithm of Sicilia and co-workers.\cite{sicilia_new_2008}
In essence, the mean energy is minimized in the subspace of the intersection space (dimension $3N-8$ for $N$ atoms) by using the gradient average projected out of the branching space (dimension 2).
The crucial difference with more usual implementations concerns the evaluation of the two vectors that span the branching space.
In particular, the derivative coupling between electronic excited states is not known routinely with TD-DFT calculations.
To circumvent this, we have evaluated numerically a pair of branching-space vectors,  $\mathbf{g}$ and $\mathbf{h}$, identified to the orthogonal eigenvectors (mass-weighted energy gradients) corresponding to the two non-zero eigenvalues of the mass-weighted Hessian of the squared energy difference calculated at the MECI geometry.\cite{gonon_applicability_2017}

\begin{figure}
  \includegraphics[width=1.00\linewidth]{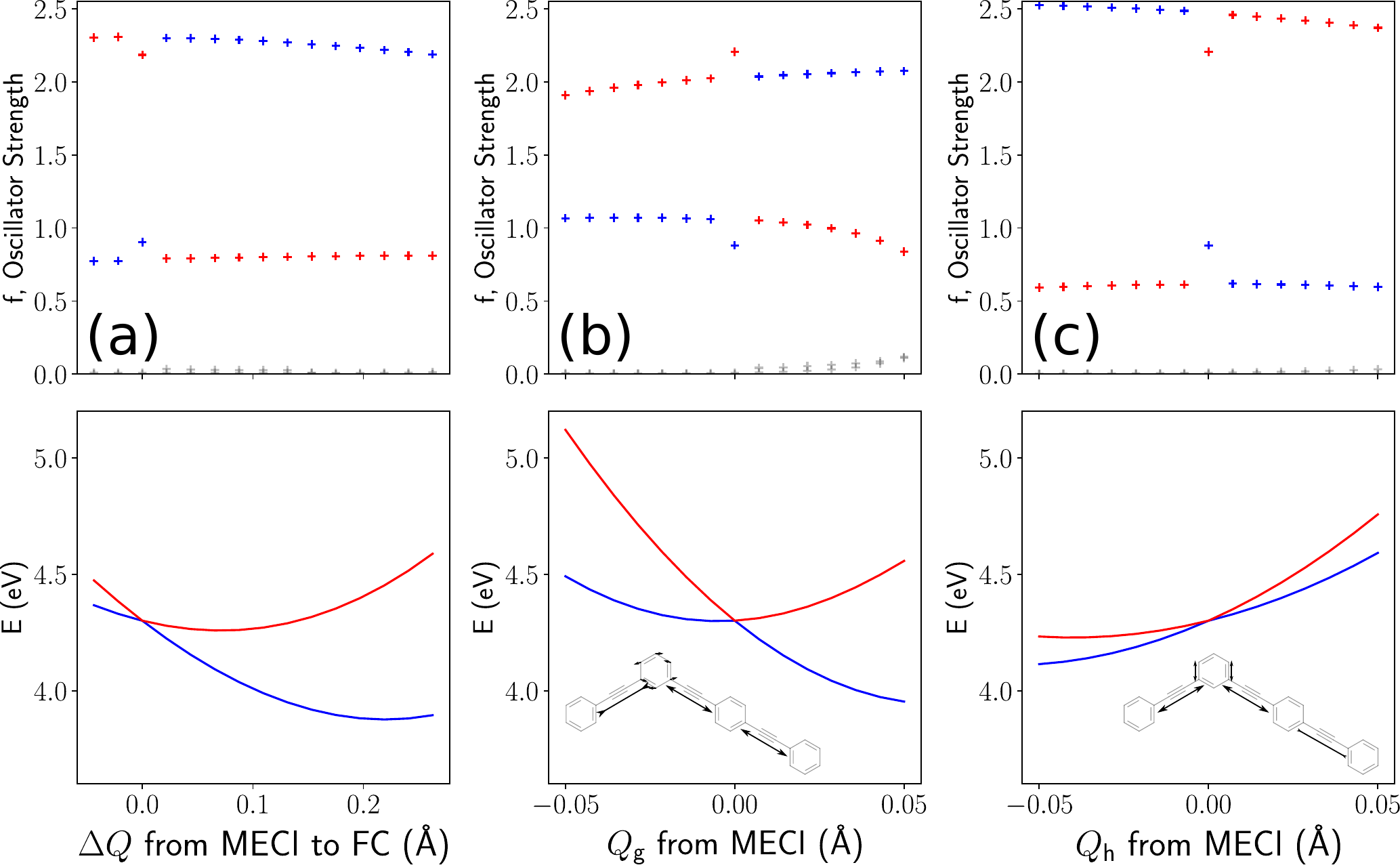}
  \caption{Oscillator strengths (top panels) and adiabatic energies (bottom panels) of the first two electronic excited states along: \textbf{(a)} the shift from the FC geometry to the MECI geometry; \textbf{(b, c)} the nuclear displacements along the branching-space vectors ($\mathbf{g}$: \textbf{(b)}; $\mathbf{h}$: \textbf{(c)}) from the MECI geometry (a shifted displacement from the MECI origin, $\Delta \mathbf{Q} = \mathbf{Q} - \mathbf{Q}_\textrm{MECI}$, was used for both profiles along either $\mathbf{g}$ or $\mathbf{h}$).
  The oscillator strengths of the next two (dark) electronic states are shown in gray for information.}
  \label{fig:hg_vectors_noRotation_schematic}
\end{figure}

At the MECI, the mean energy is $\bar{E}=\qty{4.30}{\electronvolt}$ and the energy difference is $\Delta E=E(\text{S}_2)-E(\text{S}_1)=\qty{0.0003}{\electronvolt}$.
The interpolation path between the FC geometry and the MECI geometry is given in \cref{fig:hg_vectors_noRotation_schematic} \textbf{(a)}.
The potential energy profiles from the MECI geometry along both branching-space vectors are given in \cref{fig:hg_vectors_noRotation_schematic} \textbf{(b,c)} with a simplified representation of the nuclear displacements associated to $\mathbf{g}$ and $\mathbf{h}$.
A more rigorous illustration of their associated Cartesian displacements is given in SI (fig. SI2).

Because both electronic states $\text{S}_{1}$ and $\text{S}_{2}$ have the same symmetry, A', the derivative coupling is totally-symmetric in the C\textsubscript{S} point group: in other words, we are facing here the challenging and non-usual case of a conical intersection that is not induced by symmetry.
As a result, both  branching-space vectors have the same symmetry and are only defined up to an arbitrary rotation.  However, it must be stressed that the specific pair of branching-space vectors obtained \emph{via} diagonalization of the mass-weighted Hessian of the squared energy difference at the MECI (rank-2-out-of-3$N$ matrix\cite{gonon_applicability_2017}) is not random and is determined by the fact that both mass-weighted gradients (the two nontrivial eigenvectors) have to be orthogonal and correspond to the two normal main axes of the elliptic energy cut trough the cone. We define first $\mathbf{g}$ as the short axis (largest eigenvalue) and $\mathbf{h}$ as the long axis (smaller eigenvalue), according to the ideal model of a weak coupling between localized states on the p2-branch and p3-branch.

Now, the shape of the potential energy profiles and the values of the oscillator strength along both branching-space vectors (\cref{fig:hg_vectors_noRotation_schematic}) show that the underlying pair of diabatic states that hence cross along $\mathbf{g}$ (tuning mode) and interact along $\mathbf{h}$ (coupling mode) may not be the most ideal pair of fully localized states as we have at the FC geometry.
This point will be addressed on more practical terms in \cref{sec:results_stationary}.

Considering the absence of any symmetry ``guidance'', the respective values of the two oscillator strengths at the MECI in \cref{fig:hg_vectors_noRotation_schematic} appear to be somewhat ``random'' (they -- in fact -- reflect the arbitrary gauge-invariant combination between both locally-excited  degenerate states obtained there).
Nuclear displacements along the FC-to-MECI shift tend to preserve a constant value for them, which validates the fact that a Franck-Condon-type approximation holds (almost-constant transition dipole) along this direction, together with a consistent diabatic picture for the dominant nature of both electronic states.

The moderate variation of the oscillator strengths along both branching-plane vectors is more delicate to interpret: in fact, $\mathbf{g}$ tends to couple (make oscillator strengths more identical) and $\mathbf{h}$ tends to tune (make oscillator strengths more different) both LE diabatic states with constant oscillator strengths and related to the p3 and p2 pseudofragments.

Interpolations between critical points (minima and MECI) of the PES of m23 are given in \cref{fig:interpolation_etdm} (right) along with the vertical-transition characteristics (left). They illustrate the energetics of the targeted phenomenon in the present study: EET from the photo-induced excitation from the p2 pseudofragment ($\text{S}_{2}$ adiabatic state) via a radiationless relaxation to the p3 pseudofragment ($\text{S}_{1}$ adiabatic state) (green arrows in \cref{fig:interpolation_etdm}, right).

We also computed and compared the normal modes of vibration at the minima of S\textsubscript{0}, S\textsubscript{1}, and S\textsubscript{2}. The corresponding frequencies and reduced masses of the most prominent (quinoidal and acetylenic) modes are given in \cref{tab:mode_character}.
The ones in the electronic excited states are a bit different from those in the ground state (Duschinsky rotations and second-order Jahn-Teller effect) and must be re-ordered with respect to frequencies and potentially maximize overlaps between the various sets of modes. 
As it stands, mode 119 is in particular softened in the excited state S\textsubscript{1}.
In contrast, modes 118 and 120 of the ground state overlap strongly with what are modes 119 and 138 (highest-frequency mode) of the excited state S\textsubscript{2} (before the re-assignation presented in \cref{tab:mode_character}).
This is explained by the closeness of the MinS\textsubscript{2} and MECI points (small energy gap).
Mode 138 of the excited state is in nature close to the acetylenic modes 118 and 120 of the ground state, but is frequency-displaced, because of the closeness of the intersection seam, to the C-H frequency region and allowed to mix with the corresponding modes. 
Yet, mixing the modes preserves the subset corresponding to C-H vibrations, with their typical frequencies.
\begin{table*}
  \caption{
    Frequency and reduced mass of the normal modes of vibration associated to quinoidal and acetylenic displacements in m23.
    The numbering of the normal modes is associated to the numbering of normal modes computed at MinS\textsubscript{0}.
    The normal modes at other geometries are sorted so as to satisfy the best overlap with the normal modes of the ground state geometry.
    The mass-weighted shifts between the minimum of S\textsubscript{0} and the minima of S\textsubscript{1}, S\textsubscript{2} and the MECI are also given.
    The contributions of the normal modes to the $\mathbf{g}$, $\mathbf{h}$, and gradient average $\mathbf{s}$ vectors at the MECI are given in the last columns, with the sum of the 8-dimensional model contributions.
  }
  \label{tab:mode_character}
  \begin{tabular}{l|rrrrrrrrrrrr}
    \toprule
    & \multicolumn{3}{c}{Frequency (\unit{\per\centi\meter})} & \multicolumn{3}{c}{Reduced mass (u)} & \multicolumn{3}{c}{Shift ($a_0\sqrt{m_e}$)} & \multicolumn{3}{c}{Contrib. to } \\
    Mode (S\textsubscript{0}) & MinS\textsubscript{0} & MinS\textsubscript{1} & MinS\textsubscript{2} & MinS\textsubscript{0} & MinS\textsubscript{1} & MinS\textsubscript{2} &  MinS\textsubscript{1} &  MinS\textsubscript{2} &  MECI & \%$\mathbf{g}$ & \%$\mathbf{h}$ & $\%\mathbf{s}$ \\
    \midrule
    111 & 1656 & 1616 & 1539 &  5.47 &  4.62 & 2.64  &   2.28 &    5.57 &  15.61 &   0 &  36 &  21 \\
    114 & 1682 & 1635 & 1768 &  5.87 &  4.97 & 7.54  &  -5.17 &    4.57 &   3.79 &  18 &   2 &   5 \\
    115 & 1689 & 1697 & 1675 &  5.77 &  6.09 & 5.71  &   5.76 &    2.19 &  -1.49 &   1 &   9 &   7 \\
    116 & 1693 & 1690 & 1633 &  5.87 &  5.78 & 5.22  &  -1.10 &   -2.72 &  -0.40 &   0 &   3 &   1 \\
    117 & 1699 & 1664 & 1688 &  6.04 &  5.50 & 5.82  &  -6.00 &   -1.15 &   1.82 &   3 &   6 &   8 \\
    118 & 2359 & 2267 & 2299 & 12.00 & 11.99 & 11.98 &   8.07 &    1.00 &  -0.97 &  28 &   6 &  31 \\
    119 & 2366 & 2023 & 2289 & 12.00 & 11.72 & 11.98 &   0.85 &   -2.36 &  -0.75 &  14 &  11 &   0 \\
    120 & 2367 & 2357 & 3279 & 12.00 & 12.00 & 3.03  &   0.82 &    8.02 &   7.32 &  26 &   2 &   7 \\
    Sum(8-D) & & & & & & & & & & 90 &  75 &  80 \\
    \bottomrule
\end{tabular}
\end{table*}

To characterize further the coupled PESs of m23, we also give in \cref{tab:mode_character} the mass-weighted shifts from the MinS\textsubscript{0} geometry toward the geometries of the minima of S\textsubscript{1} and S\textsubscript{2} and the MECI geometry projected onto the normal modes of MinS\textsubscript{0}.
Such projections further confirm the localization of the S\textsubscript{1} state on the p3 branch (strong shift along the p3-acetylenic mode 118) and of the S\textsubscript{2} state on the p2 branch (strong shift along the p2-acetylenic mode 120).

The shift toward the MECI, on the other hand, highlights the importance of the central-quinoidal mode 111 for describing the MECI and its seam.
Finally, the contributions of the normal modes to the gradient average and the branching-space vectors show that our eight selected modes are enough for an adequate description in the region of the MECI.
Indeed, they amount for about 90\%, 75\%, and 80\% of the contributions to $\mathbf{g}$, $\mathbf{h}$, and $\mathbf{s}$, respectively (where $\mathbf{s}$ is the gradient average at the MECI, while $\mathbf{g}$ and $\mathbf{h}$ are, as already mentioned, the branching-space vectors there).

\subsection{Linear Vibronic Coupling Without Symmetry}
We describe the PESs of the first two coupled excited states in m23 \textit{via} a linear vibronic coupling (LVC) Hamiltonian model, which corresponds to what is known as a ``diabatization by ansatz''.\cite{cederbaum_strong_1977,koppel_multimode_1984,cattarius_all_2001}
Because both electronic states have the same electronic symmetry (A') in the regions of interest, -- and because we restrict our description of vibrations to in-plane modes of A' symmetry -- the procedure is nontrivial\cite{gonon_generalized_2019} compared to symmetry-induced crossings where point-group irreducible representations can be used to discriminate how the various terms entering the potential-energy matrix may vary with coordinates of different symmetries.\cite{raab_molecular_1999}
In particular, here, there is no symmetry-labelling of the branching-space vectors at the MECI (they are only defined up to an arbitrary rotation).
As a result, for a given set of diabatic states, all normal modes can be involved in the diabatic potentials and inter-state couplings, and there is no exclusive separation of the normal modes into \textit{coupling} modes or \textit{tuning} modes.

We propose the following Franck-Condon-centered ($\mathbf{Q}=\mathbf{0}$) LVC model, with non-zero diabatic gradients and first-order inter-state coupling parameters at the origin for all the modes,
\begin{equation}
\begin{aligned}
H-T_{\text{nu}}\mathbb{1}&=
\begin{bmatrix}
E^{(1)} & 0 \\
0 & E^{(2)}
\end{bmatrix}
+
\begin{bmatrix}
\sum_i k_i^{(1)}Q_i^2 & 0 \\
0 & \sum_i k_i^{(2)}Q_i^2
\end{bmatrix}\\
&+
\begin{bmatrix}
\sum_i\kappa_i^{(1)}Q_i & 0 \\
0 & \sum_i\kappa_i^{(2)}Q_i
\end{bmatrix}
+
\begin{bmatrix}
0 & \sum_i h_i'Q_i \\
\sum_i h_i'Q_i & 0 \\
\end{bmatrix}
\quad. 
\end{aligned}
\end{equation}
The diagonal parameters gather the information on diabatic potentials: the vertical transition energies at the FC geometry $E^{(s)}$, the gradients $\kappa_n^{(s)}$ at the FC geometry (providing, together with the curvatures, an ``equivalent'' shift toward each diabatic potential minimum) and the curvatures $k^{(s)}_n$ (providing ``equivalent'' frequencies) of the diabatic potential.
The off-diagonal parameters are the components $h'_n$ of the inter-(diabatic-)state coupling vector.
In this work, $\mathbf{h}'$ is defined as a linear combination of the \textit{ab initio} branching-space vectors $(\mathbf{g},\mathbf{h})$.
Strictly speaking, this is a ``refined'' LVC model because we chose different curvatures for both diabatic states so as to account for the different bonding patterns of the two locally-excited states.
The detailed parametrization of this model will be discussed in \cref{sec:results_stationary}.

\subsection{Quantum Dynamics}

Quantum dynamics calculations were carried out using the MCTDH method as implemented in the Quantics package.\cite{meyer_multi-configurational_1990,beck_multiconfiguration_2000,worth_quantics_2020}
In this section, we briefly give the corresponding computational details and discuss the calculation of diabatic and adiabatic populations.

\subsubsection*{MCTDH Computational Details}

The single-set multiconfiguration time-dependent Hartree (MCTDH) ansatz was used here to describe the time evolution of the nuclear wavepacket of the system for two coupled electronic states.
The parameters for the primitive basis set and for the single-set SPF set are given in \cref{tab:MCTDH_parameters}.

The primitive basis set was chosen with from 9 to 11 harmonic-oscillator (Gauss-Hermite) basic functions per mode, with equilibrium geometry at zero (the FC origin), reduced mass set to one (mass-weighted coordinates), and ground-state frequencies corresponding to each normal mode of the model.

The number of single-particule functions (SPFs) for each degree of freedom was chosen so as to provide an optimal contraction of the primitive basis as regards natural weights, in particular when the lowest natural weight of each SPF occurs to be lower than \num{0.5e-3} for the whole simulation.
\begin{table}
  \caption{
    Numerical parameters for  the primitive basis [type: harmonic-oscillator (HO) Gauss-Hermite basic functions, number, frequency, and reduced mass], as well as number of single-set single-particle functions used for the propagation of the MCTDH nuclear wavepackets.
    Unit effective masses are consistent with the use of mass-weighted coordinates.
    The contraction ratio between the primitive basis and the SPF basis is approximately \num{60}.
    The final decimal places for the frequency parameters are obviously not relevant for spectroscopy, but are the numerics that we used for defining the primitive basis set with unit conversions from masses and frequencies to Gaussian widths.
  }
  \label{tab:MCTDH_parameters}
  \begin{tabular}{lccccc}
    \toprule
    Mode (S\textsubscript{0}) & Type & Size & Frequency (\unit{\per\centi\meter}) & Effective mass & \# SPF\\
    \midrule
    111 & HO   & 9    & 1655.6236 & 1.0 & 6 \\
    114 & HO   & 9    & 1681.9818 & 1.0 & 7 \\
    115 & HO   & 9    & 1689.4511 & 1.0 & 5 \\
    116 & HO   & 9    & 1693.1051 & 1.0 & 5 \\
    117 & HO   & 9    & 1699.1781 & 1.0 & 5 \\
    118 & HO   & 11   & 2359.4748 & 1.0 & 6 \\
    119 & HO   & 11   & 2366.1303 & 1.0 & 6 \\
    120 & HO   & 11   & 2367.2532 & 1.0 & 7 \\
    \bottomrule
  \end{tabular}
\end{table}

\subsubsection*{Diabatic and Adiabatic Populations}

The molecular wavepacket can be Born-expanded as
\begin{equation}
  \Ket{\Psi(\mathbf{Q,t)}} = \sum_s \Psi_s(\mathbf{Q},t) \Ket{\Phi_s;\mathbf{Q}} \quad,
\end{equation}
where the many-body electronic basis set is chosen as being orthonormal for all $\mathbf{Q}$, 
\begin{equation}
\Braket{\Phi_s;\mathbf{Q}|\Phi_r;\mathbf{Q}} = \delta_{sr} \quad,
\end{equation}
such that $\delta_{sr}$ is the Kronecker symbol. For the moment, Dirac's ``bra-ket''-notation corresponds here to an implicit integration over the electronic degrees of freedom, while the nuclear degrees of freedom (positions; $\mathbf{Q}$) have to be specified explicitly.
The corresponding density operator is defined as 
\begin{equation}
  \hat{\rho}(\mathbf{Q},t) = \Ket{\Psi(\mathbf{Q},t)}\Bra{\Psi(\mathbf{Q},t)} \quad,
\end{equation}
which expands as 
\begin{equation}
  \hat{\rho}(\mathbf{Q},t) = \sum_s \sum_r
  \Psi_s(\mathbf{Q},t) \Psi^*_r(\mathbf{Q},t)
  \Ket{\Phi_s;\mathbf{Q}}\Bra{\Phi_r;\mathbf{Q}}
  \quad.
\end{equation}
``Bra-ket''-type electronic integration brings the following discrete density matrix representation, 
\begin{equation}
  \rho_{sr}(\mathbf{Q},t) =
  \Bra{\Phi_s;\mathbf{Q}}\hat{\rho}(\mathbf{Q},t)\Ket{\Phi_r;\mathbf{Q}} =\Psi_s(\mathbf{Q},t) \Psi^*_r(\mathbf{Q},t) \quad.
\end{equation}
Now, tracing the density operator over the nuclear degrees of freedom (i.e., integrating over $\mathbf{Q}$) yields a reduced-density representation, 
\begin{equation}
\gamma_{sr}(t) = \int_\mathbf{Q} d\mathbf{Q}
  \Psi_s(\mathbf{Q},t) \Psi^*_r(\mathbf{Q},t) \quad.
\end{equation}
The diabatic electronic (integral) populations are thus defined as
\begin{equation}
  P_s(t)= \gamma_{ss}(t)
  =\int_\mathbf{Q} d\mathbf{Q}
  \vert\Psi_s(\mathbf{Q},t)\vert^2 \quad,
\end{equation}
and the diabatic electronic (integral) coherences satisfy
\begin{equation}
  C_{sr}(t)=\gamma_{sr}(t)=\gamma^*_{rs}(t)=C^*_{rs}(t) \quad,
\end{equation}
for $r \neq s$.

In what follows, Dirac's ``Bra-ket''-notation will no longer be used for integration over the electronic degrees of freedom. Hence, and unless otherwise specified, we shall now use $\Braket{\cdot|\cdot'}$ as a shorthand notation so as to refer to integration over $\mathbf{Q}$ (the nuclear degrees of freedom) only. With this in mind, we get
\begin{equation}
  P_s(t)= \Braket{\Psi_s|\Psi_s}(t) \quad,
\end{equation}
and (for $r \neq s$)
\begin{equation}
  C_{rs}(t)=\Braket{\Psi_s|\Psi_r}(t) \quad,
\end{equation}
where attention must be paid as regards index order ($r$ and $s$) and complex-conjugation.
Interestingly enough, because we are essentially dealing with a two-level electronic system, the evaluation of such magnitudes in practice simply involves to get the expectation values of Pauli operators over time.

Now, since the representation of the wavepacket is diabatic, the evaluation of adiabatic populations is to be done \textit{a posteriori}, computing upon post-processing at each time value the adiabatic transformation and integrating the resulting multi-dimensional adiabatic wavepackets over the grid, which is known to be a challenging task on the numerical front.
In this work, the adiabatic populations were obtained easily thanks to the recent implementation of the time-dependent discrete variable representation (TD-DVR) formalism in the Quantics analysis programs for single-set calculations.\cite{coonjobeeharry_mixed-quantum-classical_2022,manthe_timedependent_1996}

To validate the use of the TD-DVR integration scheme, we preliminary compared our results with usual DVR integration for multi-set calculations and observed no significant difference.

\subsubsection*{A note on Expectation Values of state-specific operators}
Let us specify some handy notations for what follows. 
Given an operator $\hat{O}$ acting on $\mathbf{Q}$ and with no off-diagonal term among the electronic manifold, the expectation value of $\hat{O}$ for the system is
\begin{equation}
  \begin{aligned}
    \Braket{\hat{O}}&=
    \frac{\Braket{\Psi|\hat{O}|\Psi}}{\Braket{\Psi|\Psi}}= \sum_s \Braket{\Psi_s|\hat{O}|\Psi_s}\\
  &= \sum_s P_s \frac{\Braket{\Psi_s|\hat{O}|\Psi_s}}{\Braket{\Psi_s|\Psi_s}}
  = \sum_s P_s \Braket{\hat{O}}_s \quad,
  \end{aligned}
\end{equation}
assuming $\Braket{\Psi|\Psi}=1$ and where the electronic populations (weights) are $P_s=\Braket{\Psi_s|\Psi_s}$.

The expectation value for the whole system can thus be interpreted as a population-weighted sum of state-specific expectation values $\Braket{\hat{O}}_s$ (with state-occurence probability weights -- populations -- $P_s$). In other words, two types of state-specific expectation values can be discussed:
\begin{itemize}
  \item
  normalized (state-specific) expectation values:
  \begin{equation}
  \braket{\hat{O}}_s=\frac{\Braket{\Psi_s|\hat{O}|\Psi_s}}{\Braket{\Psi_s|\Psi_s}}
  \quad,
  \end{equation}
  which are ``intensive'' expectation values in the sense that they cannot be added together;
  \item
  population-weighted (for any state-contribution) expectation values:
  \begin{equation}
  \Braket{\Psi_s|\hat{O}|\Psi_s} = P_s \Braket{\hat{O}}_s
  \quad,
  \end{equation}
  which are ``extensive'' expectation values in the sense that they can be added together to assemble the full expectation value upon adding all state-contributions.
\end{itemize}
In single-set calculations, the population-weighted expectation values are directly obtained, and must be normalized to access the state-specific expectation values.
Normalized state-specific expectation values are required when studying the system in some given electronic state during a phenomenon that involves population transfer.

Conceptually speaking, state-specific expectation values correspond to what could be measured ideally (for example, a bond length, assuming specific knowledge of the electronic state). In contrast, a state-contribution expectation value is already weighted by the probability of observing the state and tells us how much it partially contributes to a global one (for example to the total energy).

\section{Potential Energy Surfaces and Stationary Properties}
\label{sec:results_stationary}
\subsection{Parametrization of a Vibronic Coupling model without symmetry}
The LVC model used in the present work was parameterized numerically so as to fit \textit{ab initio} energy and derivative data calculated at various geometry points.

The parameters for the off-diagonal term, the components of the $\mathbf{h}'$-vector, are obtained upon projection of the approximate branching-space vectors at the MECI point on the normal modes.
The parameters for the diagonal terms are the vertical transition energies, gradients at the FC geometry, and curvatures, obtained upon adjusting adiabatic energies to \textit{ab initio energies on} rigid scans along normal-mode displacements for the already fixed $\mathbf{h}'$.
This way of choosing the parameters is a compromise between interpolation and extrapolation.

As we want our model to be accurate not only in the Franck-Condon region, we did not use local \textit{ab initio} derivatives calculated there but preferred a more global approach based on fitting the gradients and curvatures to rigid scans along displacements associated to the normal modes of vibration of the ground state.
Because the energy gap is significant around the origin, the effect of the diabatic coupling on the adiabatic curvatures is small there and many ``quasi-harmonic adiabatic solutions'' can be obtained when optimizing freely both the diabatic curvatures and the coupling vector.
For this reason, we decided to leave the $\mathbf{h}'$-vector parameters out of the fitting procedure, and fix them from the onset as an \textit{ab initio} input.

A reasonable assumption for choosing $\mathbf{h}'$ in advance is to consider that it does not change from the MECI point to the FC geometry because the locally-excited diabatic states are essentially the same here and there. However, rotational freedom between both branching-space vectors $(\mathbf{g},\mathbf{h})$ (due to gauge invariance between degenerate electronic states at the MECI) does not guarantee this directly, unless a specific rotation is further defined so as to enforce maximal diabatic correspondence with the FC geometry (see, for example, Ref \cite{gonon_generalized_2019}). 

To this end, we define a rotation $(\mathbf{g},\mathbf{h})\rightarrow (\mathbf{g}',\mathbf{h}')$ parameterized with
\begin{equation}
\begin{aligned}
    \tan{\theta}=-\frac{\mathbf{h}\cdot\Delta\mathbf{Q}_X}{\mathbf{g}\cdot\Delta\mathbf{Q}_X}
\end{aligned}
\quad,
\end{equation}
such that the $\mathbf{h}'$-vector becomes orthogonal to the shift $\Delta \mathbf{Q}_X$ between the FC geometry and the MECI geometry.
This procedure is consistent with ensuring that diabatic and adiabatic states coincide at the reference FC geometry, which occurs to be convenient for further interpretation of diabatic and adiabatic populations.

As already pointed out, setting an LVC model in the absence of symmetry is a nontrivial task and requires to make decisions as regards the nature of the underlying diabatic states. Our choice of parametrization should describe the best of two worlds: accurate, fitted, parameters at the FC geometry, in particular gradients; and accurate, identified, coupling parameters from a relevant conical intersection point (the MECI).
The description of the Franck-Condon region is crucial for the initialization and the early dynamics of the system, while the description of the MECI is important for having a reliable estimate of the EET kinetics.

From a practical point of view, we fit the linear diabatic parameters of the profiles one by one.
Thus, for one rigid scan, we have a set of displaced geometries $\{\mathbf{Q}_{n,l}\}_{l\in \text{scan}}=\{(0,\cdots,Q_l,\cdots,0)\}_{l\in\text{scan}}$, each along only one normal mode.
The parameters of the LVC Hamiltonian are thus obtained by fitting its eigenvalues (adiabatic energies) to the \textit{ab initio energies} along each rigid scan.
This is implemented as a minimization of the least-square function for normal mode $n$,
\begin{equation}
  \begin{aligned}
  &L(\kappa_n^{(1)},\kappa_n^{(2)},k_n^{(1)},k_n^{(2)}) =\\
  &\sum_{s=1,2}
  \sum_{l\in\text{scan}}\left(V^{(s)}[\kappa_n^{(1)},\kappa_n^{(2)},k_n^{(1)},k_n^{(2)},h'_{n}](\mathbf{Q}_{n,l}) - E^{(s)}(\mathbf{Q}_{n,l})\right)^{2}
  \end{aligned}
  \quad,
\end{equation}
where $V^{(s)}[\kappa_n^{(1)},\kappa_n^{(2)},k_n^{(1)},k_n^{(2)},h'_{n}](\mathbf{Q}_{n,l})$ are the eigenvalues of the LVC Hamiltonian, and $E^{(s)}(\mathbf{Q}_{n,l})$ the \textit{ab initio} energies at the displaced geometry $\mathbf{Q}_{n,l}$.
Note again that the parameters $h'_{n}$ are not optimized, and given as an input to the fitting procedure.

The resulting parameters are given in \cref{tab:parameters_equivalent} in the form of ``equivalent'' magnitudes in terms of frequencies and shifts.
The numerical values associated to $\kappa_n^{(1)},\kappa_n^{(2)},k_n^{(1)},k_n^{(2)},h'_{n}$ for implementing the operators used in quantum dynamics calculations are given in mass-weighted atomic units in the SI (Table SI1).
\begin{table}
  \caption{
    Equivalent quantities for LVC parameters obtained upon fitting \textit{ab initio} calculations.
    Parameters associated with the harmonic (second-order) expansion ($k_{i}^{(n)}$) are provided in terms of the associated frequency,  $\omega_{i}^{(n)}$.
    Parameters associated with the first-order expansion ($h'_{i}$ and $\kappa_{i}^{(k)}$) are provided also in terms of their  characteristic induced geometry shifts,  $d_{i}'^{(12)}=-\frac{2h_{i}'}{k_{i}^{(1)}+k_{i}^{(2)}}$ and $d_{i}^{(k)}=-\frac{\kappa_{i}^{(k)}}{k_{i}^{(k)}}$.
    The corresponding LVC parameters in mass-weighted atomic units are gathered in Table SI1.
  }
  \label{tab:parameters_equivalent}
  \begin{tabular}{lccrrr}
    \toprule
    Parameters & \multicolumn{2}{c}{Curvatures} & \multicolumn{3}{c}{Gradients} \\
    & $k_{i}^{(1)}$ & $k_{i}^{(2)}$ & $\kappa_{i}^{(1)}$ & $\kappa_{i}^{(2)}$ & $h'_i$ \\
    Equivalent & \multicolumn{2}{c}{Frequencies (\unit{\per\centi\meter})} & \multicolumn{3}{c}{Shifts ($a_{0}\sqrt{m_{e}}$)} \\
    & $\omega_{i}^{(1)}$ & $\omega_{i}^{(2)}$ & $d_{i}^{(1)}$ & $d_{i}^{(2)}$ & $d'^{(12)}_{i}$ \\
    Mode & & & & & \\
    \midrule
    111   &     1650  &    1554 &  2.54 &  6.06 & -1.92 \\
    114   &     1666  &    1437 & -5.28 &  7.29 &  2.63 \\
    115   &     1648  &    1682 &  5.97 &  1.81 &  0.58 \\
    116   &     1684  &    1599 & -1.00 & -3.61 & -0.71 \\
    117   &     1644  &    1692 & -6.15 & -0.55 & -0.16 \\
    118   &     2345  &    2301 &  8.80 &  0.98 & -0.62 \\
    119   &     2103  &    2157 &  1.16 & -2.38 & -1.46 \\
    120   &     2342  &    2108 &  1.62 & 11.10 &  1.46 \\
    \bottomrule
  \end{tabular}
\end{table}

The contribution of each mode to the inter-state coupling, the diabatic gradient half-difference (further denoted gradient difference for simplicity), and the diabatic gradient average within our reduced 8-dimensional (8-D) subspace are gathered in \cref{tab:mode_contributions_model}.
Such a decomposition highlights the importance of the acetylenic modes involving the p2 branch, 119 and 120.
The latter two account together for more than the third of each of the three vectors within the reduced 8-D model.
The third acetylenic mode 118, localized on the p3 branch, contributes mostly to the gradient difference (tuning) and gradient average (tilting).
The quinoidal modes of the central \textit{meta}-substituted ring are also highlighted as strongly coupling (for 111, stretching mode) and strongly coupling and tuning (for 114, rock-bending mode).
\begin{table}
  \caption{
    Contribution (in percent) of each mode to the coupling vector $\mathbf{h}'$, the gradient difference vector $\mathbf{g}'$, and the gradient average vector $\mathbf{s}$.
  }
  \label{tab:mode_contributions_model}
  \begin{tabular}{lrrr}
    \toprule
    Mode & $\% \mathbf{h}'$ & $\% \mathbf{g}'$ & $\%\mathbf{s}$ \\
    \midrule
    111 & 14 &  1 &  6 \\
    114 & 19 & 17 &  0 \\
    115 &  2 &  2 &  6 \\
    116 &  2 &  1 &  2 \\
    117 &  0 &  5 &  4 \\
    118 &  7 & 37 & 37 \\
    119 & 29 &  5 &  1 \\
    120 & 27 & 32 & 44 \\
    \bottomrule
  \end{tabular}
\end{table}
\subsection{Validation and discussion of the model}
The energy profiles from rigid scans (\textit{ab initio} inputs) and from the optimized LVC model for adiabatic PESs (eigenvalues) are given in the SI and show almost perfect agreement (fig. SI3).
The diabatic potentials of the model are also given and illustrate the effect of $\mathbf{h}'$, expected to be small in the Franck-Condon region, as already pointed out.

The optimized critical points within the LVC 8-D model are in qualitative agreement with the critical points obtained from full-dimensional optimization in the \textit{ab initio} PESs (see \cref{tab:critical_points} for comparison).
In particular, the minima of the first and second adiabatic states in the model are less than \qty{0.1}{\electronvolt} off in energy from optimized critical points.
Their shifts with respect to the FC geometry along the eight normal modes included in the model are in good qualitative agreement (see Table SI2 for comparison).
The same is found for the MECI (where the ``apparent'' MECI of the model was taken as the lowest-average-energy crossing point optimized with an energy difference of $\qty{0.0001}{\hartree}=\qty{0.0027}{\electronvolt}$).

This clearly shows that our 8-D model is a perfectly acceptable trade-off between size and accuracy, since relaxing the remaining 130 frozen modes out of a total of 138 only lowers typical energies by about \qty{0.1}{\electronvolt}.

Finally, we can consider that the good agreement between the gradient difference at the FC geometry and at the MECI is an encouraging \textit{a posteriori} proof of the validity of the linear vibronic coupling model.
We compare in \cref{tab:GD_PrimeGVector} the gradient difference from the fitting procedure at the FC geometry and $\mathbf{g}'$, the branching-space counterpart of $\mathbf{h}'$ at the MECI.
We find that with the previously described rotation, vectors $\mathbf{g}'(\text{MECI})$ and $\text{\textbf{GD}}(\text{FC})$ are aligned indeed, thus showing little energy-difference path ``curving'' (which would then be due to strong differential second-order effects between both states).

\begin{table}
  \caption{
    Halved-gradient difference from the fitting procedure at the FC geometry $\text{\textbf{GD}}(\text{FC})$ and second branching-space vector $\mathbf{g}'$, counterpart of the first branching-space vector $\mathbf{h}'$ orthogonalized to the shift $\Delta \mathbf{Q}_X$ between FC geometry and MECI.
    Both quantities are given in thousandth of energy gradients given in mass-weighted atomic units ($\frac{\num{e-3}E_h}{a_{0}\sqrt{m_{e}}}$)
  }
  \label{tab:GD_PrimeGVector}
  \begin{tabular}{lrrr}
    \toprule
    Mode & $\textbf{GD}(\text{FCP})$ & $\mathbf{g}'(\text{MECI})$ & $\vert\textbf{GD}(\text{FCP}) - \mathbf{g}'(\text{MECI})\vert$ \\
    \midrule
    111 & -0.080 & -0.083 & 0.003 \\
    114 & -0.308 & -0.332 & 0.024 \\
    115 &  0.115 &  0.091 & 0.024 \\
    116 &  0.066 &  0.020 & 0.046 \\
    117 & -0.156 & -0.142 & 0.024 \\
    118 &  0.448 &  0.443 & 0.005 \\
    119 &  0.168 &  0.290 & 0.122 \\
    120 & -0.420 & -0.402 & 0.018 \\
    \bottomrule
  \end{tabular}
\end{table}

\subsection{Initial state preparations and stationary spectra}
In the simulations that follow, two different initial states are considered.
Assuming a sudden excitation from the electronic ground state minimum we choose as initial states the 8-D Gaussian wavefunctions corresponding to the ground vibrational state, projected onto either the first or second diabatic electronic state, respectively.

As already pointed out, the specific rotation used to specify the two branching-space vectors ensures that the diabatic states match as much as possible the adiabatic states at the FC geometry.
We can check this by the adiabatic populations at $t = \qty{0}{\femto\second}$ and see that the initial state on diabatic state D\textsubscript{1} is almost a 95\%:5\% mixture of the adiabatic states, while the initial state on diabatic state D\textsubscript{2} is the opposite 5\%:95\% mixture of the adiabatic states.
The remaining non-zero population on the untargeted state is due to the dispersion of the initial wavepacket in the direction of the inter-state coupling vector ($\mathbf{h}'$).

Before studying the real-time dynamics of the system upon excitation, we shall start with discussing energy-resolved stationary properties.
Absorption and emission power spectra were obtained from our simulations according to the typical procedure (Fourier transform of the autocorrelation function) used for example in seminal works by Raab or Cattarius and co-workers for pyrazine or the butatriene cation.\cite{raab_molecular_1999,cattarius_all_2001}
We also applied this method recently to compute the vibronic absorption and emission spectra of a symmetrical (2+2)-ring PPE, m22.\cite{galiana_unusual_2023}

According to the initial state preparations, we have two \emph{numerical experiments}, exciting diabatic states D\textsubscript{1} or D\textsubscript{2} localized on the p3 or p2 branch, respectively.
From these, we can access different contributions to the absorption spectrum.
The emission spectrum is then simply obtained by propagating the vibronic ground state of the excited state manifold in the electronic ground state PES.
The vibronic ground state in the excited state manifold is unambiguous, as its decomposition is almost exclusively on the diabatic state D\textsubscript{1} (population $P(\text{D}_1)=0.99$), and is mostly displaced along modes involving the p3 pseudo-fragment.

\begin{figure}
  \includegraphics[width=0.95\linewidth]{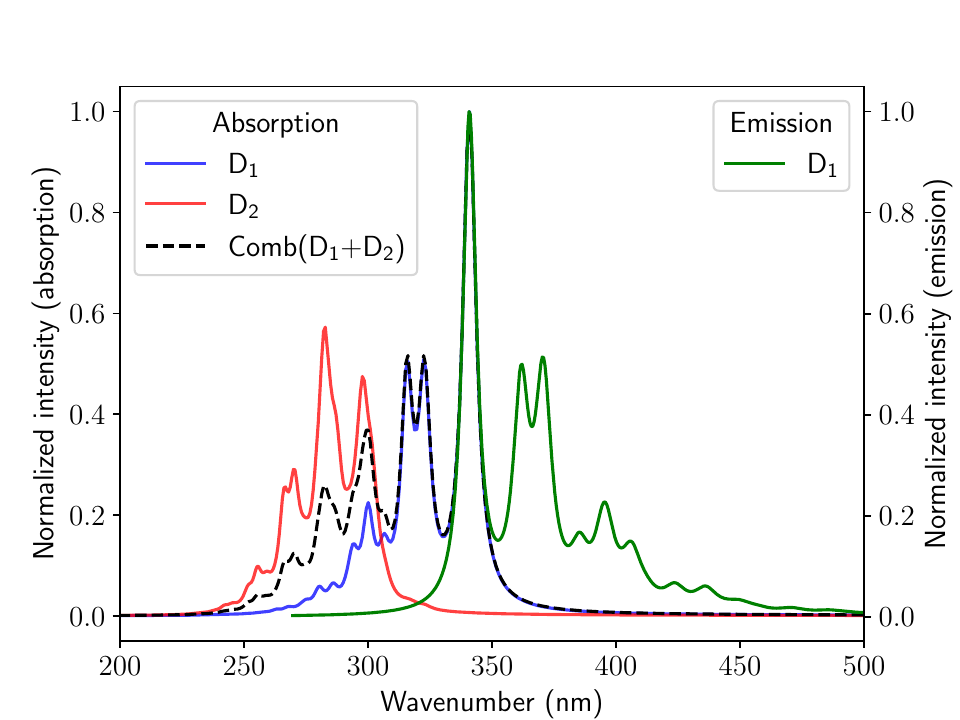}
  \caption{Power spectra contributions obtained from Fourier transform of the relevant autocorrelation functions.
    Contribution to absorption from the excitation of D\textsubscript{1}, D\textsubscript{2}, or both: plain blue, red lines, and dashed black line, respectively.
    Emission from the vibronic ground state in the (D\textsubscript{1}, D\textsubscript{2}) manifold: plain green line.
    The spectra are realistically broadened by using the damping time $\tau=\qty{19}{\femto\second}$.
  }
    \label{fig:Quantics_abs_emi}
\end{figure}

\begin{figure}
  \includegraphics[width=0.95\linewidth]{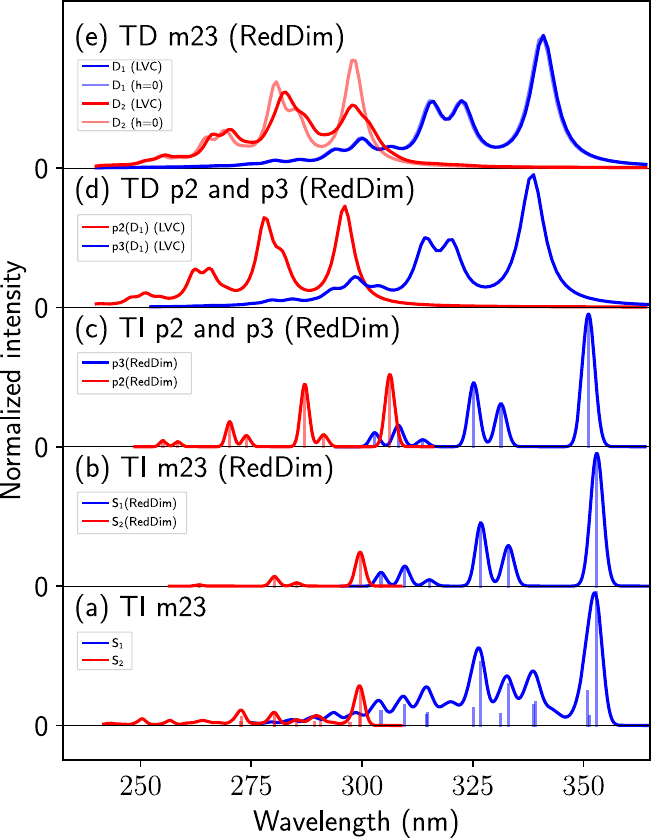}
  \caption{
    Calculated UV-visible spectra using a time-independent (TI) method giving Franck-Condon factors within the harmonic and Born-Oppenheimer approximations (a, b, c), and calculated power spectra using a time-dependent (TD) method beyond the Born-Oppenheimer approximation (d, e).
    \textbf{(a)} Full dimensional TI spectra between S\textsubscript{0} and S\textsubscript{1}/S\textsubscript{2} (blue/red lines) of m23.
    \textbf{(b)} Reduced dimensional TI spectra between S\textsubscript{0} and S\textsubscript{1}/S\textsubscript{2} (blue/red lines) of m23, selecting only transitions involving modes of the 8-D model.
    \textbf{(c)} Reduced dimensional TI spectra between S\textsubscript{0} and S\textsubscript{1} in p3/p2 fragments (blue/red lines).
    \textbf{(d)} Reduced dimensional TD power spectra between S\textsubscript{0} and D\textsubscript{1} in p3/p2 fragments (blue/red lines).
    \textbf{(e)} Reduced dimensional TD spectra between S\textsubscript{0} and D\textsubscript{1}/D\textsubscript{2} (blue/red lines) of m23 with and without coupling (plain and transparent lines).
  }
  \label{fig:all_absorption_spectra}
\end{figure}

We show absorption and emission power spectra in \cref{fig:Quantics_abs_emi}.
There are two clearly distinct contributions to the absorption spectrum.
The one from excitation on diabatic state $\text{D}_{2} \simeq \text{p2} \simeq \text{S}_{2}$ yields the most energetic transitions and the one from excitation on $\text{D}_{1} \simeq \text{p3} \simeq \text{S}_{1}$ yields the less energetic transitions.
As expected from PPE dendrimer building blocks, they match the vibronic progressions in the absorption spectra of the isolated p2 and p3 pseudo-fragments, respectively, with the peculiarity for $\text{D}_{2}$ that the band origin is not the most intense transition (see discussion below and \cref{fig:all_absorption_spectra}).

Assuming that Kasha's rule holds, the emission power spectrum is obtained upon projecting to the ground state the lowest vibronic state within the excited-state manifold, which is almost exclusively expanded on $\text{D}_{1} \simeq \text{p3} \simeq \text{S}_{1}$. We thus observe typical mirror-image spectra between emission and the first absorption band.
Note that the situation is different for the emission of the symmetrically (2+2)-ring PPE, m22 for which
the lowest vibronic state is a significant mixture of the two excited electronic states.\cite{galiana_unusual_2023}

For an estimation of what should be the total absorption spectrum of m23, we also give the sum of both contributions with the oscillator strengths of both excited states (at the FC geometry) as weighting factors (black line in \cref{fig:Quantics_abs_emi}).

For further discussion of the power spectra, we compare the quantum dynamics results to absorption spectra simulated in the time-independent (TI) framework, as implemented in the Gaussian package.\cite{santoro_effective_2007,santoro_effective_2008,barone_fully_2009}
The absorption spectra of m23, p2, and p3 are computed, with full-dimensionality or reduced-dimensionality (RedDim) including the normal modes from the 8-D (quinoinal and acetylenic) model.

Comparing the power spectra for absorption toward the two coupled diabatic states in the LVC model and the same without coupling (LVC with $\mathbf{h}'=\mathbf{0}$) in \cref{fig:all_absorption_spectra}, \textbf{(e)} exhibits some difference around \qty{300}{\nano\meter}, which is a signature of the role of nonadiabatic effects.

Indeed, the absorption toward diabatic state D\textsubscript{2} has different maxima depending on the presence or absence of coupling in the LVC model.
This can be explained by the closeness between the minimum of the second diabatic potential and the crossing, which induces significant spatial delocalization of the vibrational component over the two wells, hence a smaller overlap with the ground state.
On the contrary, the absorption toward diabatic state D\textsubscript{1} is almost unchanged by turning off the coupling, which indicates a weak anharmonicity for the lowest eigenvalues of the first adiabatic excited state.

According to the pseudo-fragmentation picture, we confirm that the spectra for the uncoupled model are analogous to the spectra from isolated p2 and p3 minimal models, computed with the same TD methodology (\cref{fig:all_absorption_spectra} \textbf{(d)}).
A shift of about \qty{2}{\nano\meter} is observed between the absorption of S\textsubscript{1} in m23 and the absorption of S\textsubscript{1} in p3, consistent with slightly different vertical transition energies (\qty{3.877}{\electronvolt} and \qty{3.896}{\electronvolt}, respectively, yielding a gap of \qty{1.6}{\nano\meter} in the region of the 0--0 (band origin) transition).
A similar shift is obtained between the absorption of S\textsubscript{2} in m23 and the absorption of S\textsubscript{1} in p2 (vertical transition energies \qty{4.452}{\electronvolt} and \qty{4.475}{\electronvolt}, respectively).

Let us now compare TD and TI absorption spectra. In the latter case, the oscillator strengths of the associated transitions are taken into account and the harmonic approximation is used.
We give the full-dimensional Franck-Condon spectra for state S\textsubscript{1} and S\textsubscript{2} in \cref{fig:all_absorption_spectra} \textbf{(a)} and the same but ``trimmed'' spectra obtained by selecting only the transitions involving the 8 modes of the model \textbf{(b)}.
Both spectra are analogous as the main transitions involve the selected quinoidal and acetylenic modes; the missing transitions in \textbf{(b)} involve essentially other anti-quinoidal rock-bending modes or triangular ones.\cite{ho_vibronic_2017}

We also give the corresponding TI absorption spectra for the isolated p2 and p3 fragments, \textbf{(c)}.

Among all types of absorption spectra, we can notice some variation around \qty{340}-\qty{350}{\nano\meter} of the spectral position of the lowest band origin (0--0 transition) corresponding to the first excited state of either m23 or its p3 isolated pseudo-fragment. 
The most intense peak occurs at: \qty{353}{\nano\meter}, \qty{351}{\nano\meter}, \qty{338}{\nano\meter} and \qty{341}{\nano\meter} in spectra \textbf{(b,c,d,e)}, respectively.

The slight shift of about \qty{2}{\nano\meter} is only due to the aforementioned small difference in transition energies to the first excited state of either m23 or the isolated p3 pseudo-fragment.

The explanation for the significant shift of about \qty{12}{\nano\meter} (typically between the full-dimensional TI spectrum \textbf{(a)} and the reduced-dimensional TD spectrum \textbf{(e)} of m23) is twofold and due to the reduction of dimensionality.
First, the energy difference for the optimized minimum of S\textsubscript{1} either \textit{ab initio} or within the LVC model is \qty{0.06}{\electronvolt}, which accounts for a shift of \qty{6}{\nano\meter}.
Then, the value of the difference of zero-point energies (ZPE) between S\textsubscript{0} and S\textsubscript{1} is not the same in the full-dimensional and reduced-dimensional systems.
With all (138) normal modes, $\text{ZPE}(\text{S}_1)=\qty{82401}{\per\centi\meter}$ and $\text{ZPE}(\text{S}_0)=\qty{83156}{\per\centi\meter}$ for a $\Delta \text{ZPE} = \qty{-755}{\per\centi\meter}$.
With the selected (8) normal modes, $\text{ZPE}(\text{S}_1)=\qty{6641}{\per\centi\meter}$ and $\text{ZPE}(\text{S}_0)=\qty{6909}{\per\centi\meter}$ for a $\Delta \text{ZPE} = \qty{-268}{\per\centi\meter}$.
The 0--0 transition in the reduced model is thus $\qty{487}{\per\centi\meter}=\qty{0.06}{\electronvolt}$ lower than in the full-dimensional system, accounting for another \qty{6}{\nano\meter} shift with respect to the 0--0 transition of the full-dimensional system.
The calculated absorption spectra toward the first excited state in the different models are reproduced in SI (fig. SI4), accounting for the energy shift related to harmonic ZPE differences for TD spectra.

Finally, let us stress that the previous steady-state spectroscopy results also serve as a validation for the choice of selected normal modes for our reduced model.
\textit{A priori} identifying the optically active modes gives a minimal set to be included if one wants a correct description of the initial state upon sudden excitation.
\textit{A posteriori} calculation of the stationary spectra validates the description of the PES around the FC geometry (and the MECI if nonadiabatic effects are observed).

The criteria for validating our model are based on stationary properties only, such as contributions to the branching plane or contributions to the stationary vibronic spectra.
We note that more involved ways of selecting the electronic states and normal modes of vibration have been proposed, for instance with a systematic self-consistent optimization procedure through both trajectory-based and quantum dynamics simulations
.\cite{gomez_surface_2019}
\section{Time-Resolved Study of Excitation Energy Transfer}
\label{sec:results_dynamics}
\subsection{Population transfer kinetics}
\begin{figure}
  \includegraphics[width=0.85\linewidth]{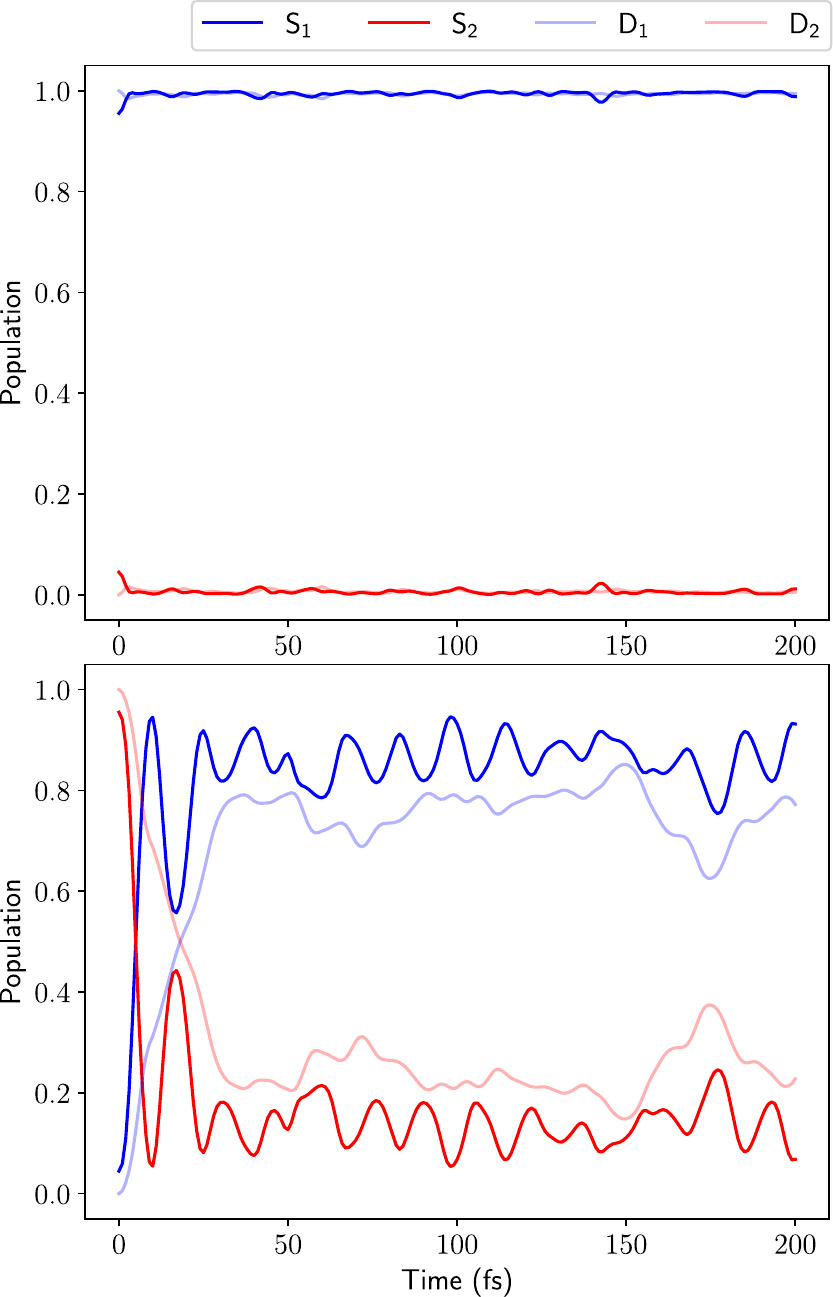}
  \caption{
    Time evolution of the populations of the first two adiabatic and diabatic states for different initial states (top: D\textsubscript{1}, bottom: D\textsubscript{2}) for the first \qty{200}{\femto\second} of simulation.
  }
  \label{fig:populations_200}
\end{figure}

We now discuss the real-time dynamics of the 2-state 8-D model of m23.
As already mentioned, we consider two limiting situations: sudden excitation either to the LE state on the p2 branch (for simulating the existence of direct EET) or to the LE state on the p3 branch (for confirming the absence of reverse EET).
For both situations, we illustrate this with the time evolution of the diabatic and adiabatic populations (\cref{fig:populations_200}).

Note that we are primarily interested here in EET from S\textsubscript{2} to S\textsubscript{1} and did not consider any further quenching through internal conversion from S\textsubscript{1} to S\textsubscript{0}; in other words, S\textsubscript{1} is assumed to be a fluorescent species whereby S\textsubscript{0}-S\textsubscript{1} couplings are negligible and the energy difference $\Delta E(\text{S}_{0}-\text{S}_{1})$ is large enough.

As expected, our result show that initial excitation to the first diabatic state (mostly S\textsubscript{1} and an LE-state on p3) has virtually no effect as regards population transfer. On the contrary, excitation to the second diabatic state (mostly S\textsubscript{2} and a LE-state on p2) exhibits quite efficient and ultrafast EET from the short to the long branch, illustrated by the transfer of population (internal conversion) to the lower state.
Indeed, we observe 90\% of population transfer from S\textsubscript{2} to S\textsubscript{1} in about \qty{25}{\femto\second}, with small fluctuations afterwards to the end of the simulation time.

Interestingly enough, we observe a difference in the time evolution of the adiabatic and diabatic populations.
Let us focus on the populations of the second state, S\textsubscript{2} and D\textsubscript{2}, the decaying state.
We observe a monotonous transfer of population from the diabatic state D\textsubscript{2} $\simeq$ p2, while the corresponding adiabatic state decays faster and populates again before decaying, with two large-amplitude oscillations in the first \qty{25}{\femto\second}.
After this time, the adiabatic population stabilizes in a decayed regime with a population below $20\%$.
We can interpret these variations in the adiabatic population as a double-crossing of the conical intersection seam.

At $t=0$, the simulation is initialized on D\textsubscript{2}, mainly S\textsubscript{2}, and quickly follows the gradient of D\textsubscript{2} toward the crossing region (in less than \qty{10}{\femto\second}).
During this early dynamics, the system mainly stays in D\textsubscript{2} but crosses the CI seam, transferring from S\textsubscript{2} to S\textsubscript{1}.
At this point, the energy difference is small enough to induce an efficient nonadiabatic coupling, allowing a fast transfer from D\textsubscript{2} to D\textsubscript{1} but the system is allowed to transfer back into S\textsubscript{2} for the second oscillation, re-crossing the CI seam.

\begin{figure*}
    \centering
    \includegraphics[width=0.9\textwidth]{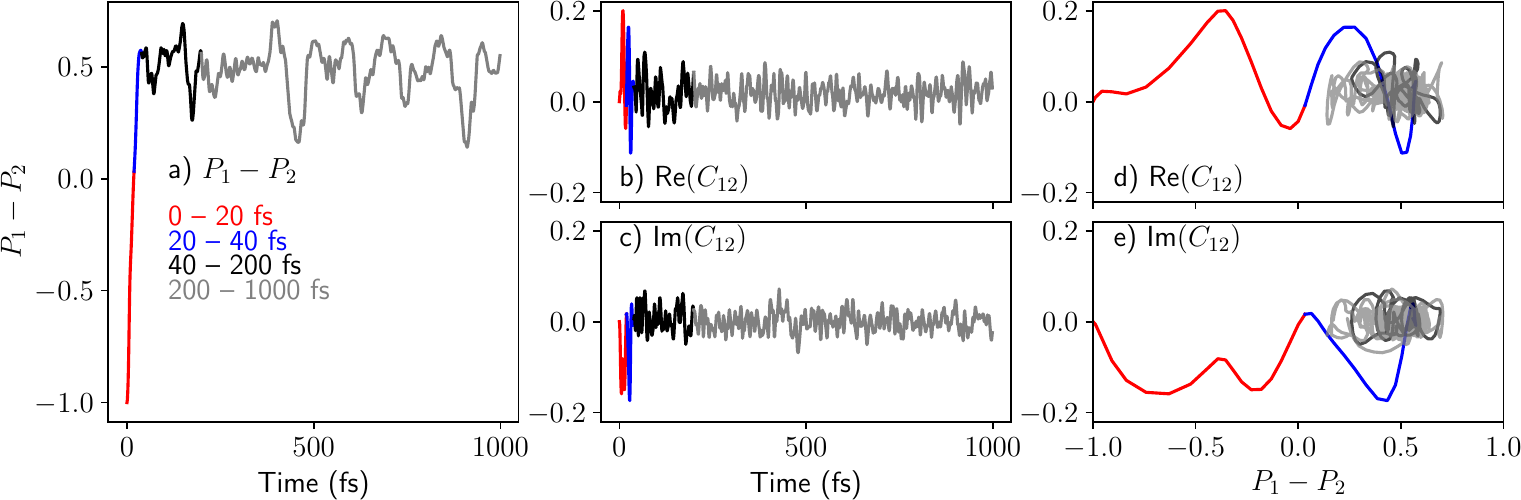}
    \caption{Time evolution of \textbf{(a)} the diabatic population difference, \textbf{(b)} the real part of the diabatic coherence, and \textbf{(c)} the imaginary part of the diabatic coherence. 
    Real and imaginary parts of the diabatic coherence as functions of the diabatic population difference are given in \textbf{(d)} and \textbf{(e)}, respectively.
    Colors correspond to different time windows.
    }
    \label{fig:C12_DeltaP}
\end{figure*}

From now on, we consider only diabatic states properties (populations, coherence, expectations values) and not adiabatic states properties.
We give the time evolution of the population difference $P_1-P_2$ and of the coherence (both real and imaginary parts) between the diabatic states up to \qty{1000}{\femto\second} in \cref{fig:C12_DeltaP} \textbf{(a)}, \textbf{(b)} and \textbf{(c)}.
The coherence between the diabatic electronic states remains significant all along the EET process, with significant maxima ($\left|C_{12}\right|=\num{0.20}$).
After the transfer, the system remains mostly in D\textsubscript{1} still exchanging population with D\textsubscript{2}.
We also illutrate the correlation between coherence and population difference in \cref{fig:C12_DeltaP} \textbf{(d)} and \textbf{(e)}.
We interpret this result as a signature of the nonadiabatic EET, with three distinct regimes.
The first regime (red line, \qtyrange{0}{20}{\femto\second}) is the first part of the population inversion, leading to equalization of the diabatic populations.
The second regime (blue line, \qtyrange{20}{40}{\femto\second}) is the completion of the population inversion, with the system ending up trapped in the lowest-lying diabatic excite state D\textsubscript{1}.
Finally, the late dynamics (black line, \qtyrange{40}{200}{\femto\second} and gray line \qtyrange{200}{1000}{\femto\second}) corresponds to oscillations in the region close to equilibrium for the system.
Future works will address a similar analysis for the adiabatic populations and coherence and further discuss the possible role of coherence in the process of nonadiabatic EET.

In the following, we characterize the EET process through the discussion of different observable related to the changes of geometry in the molecule. 
EET will be analyzed in detail within the first \qty{200}{\femto\second} of the propagation time by comparing both electronic states properties.
Long-term dynamics (up to \qty{1000}{\femto\second}) will then be shortly discussed.
\subsection{Mode-specific analysis of EET}
\begin{figure*}
  \includegraphics[width=0.48\linewidth]{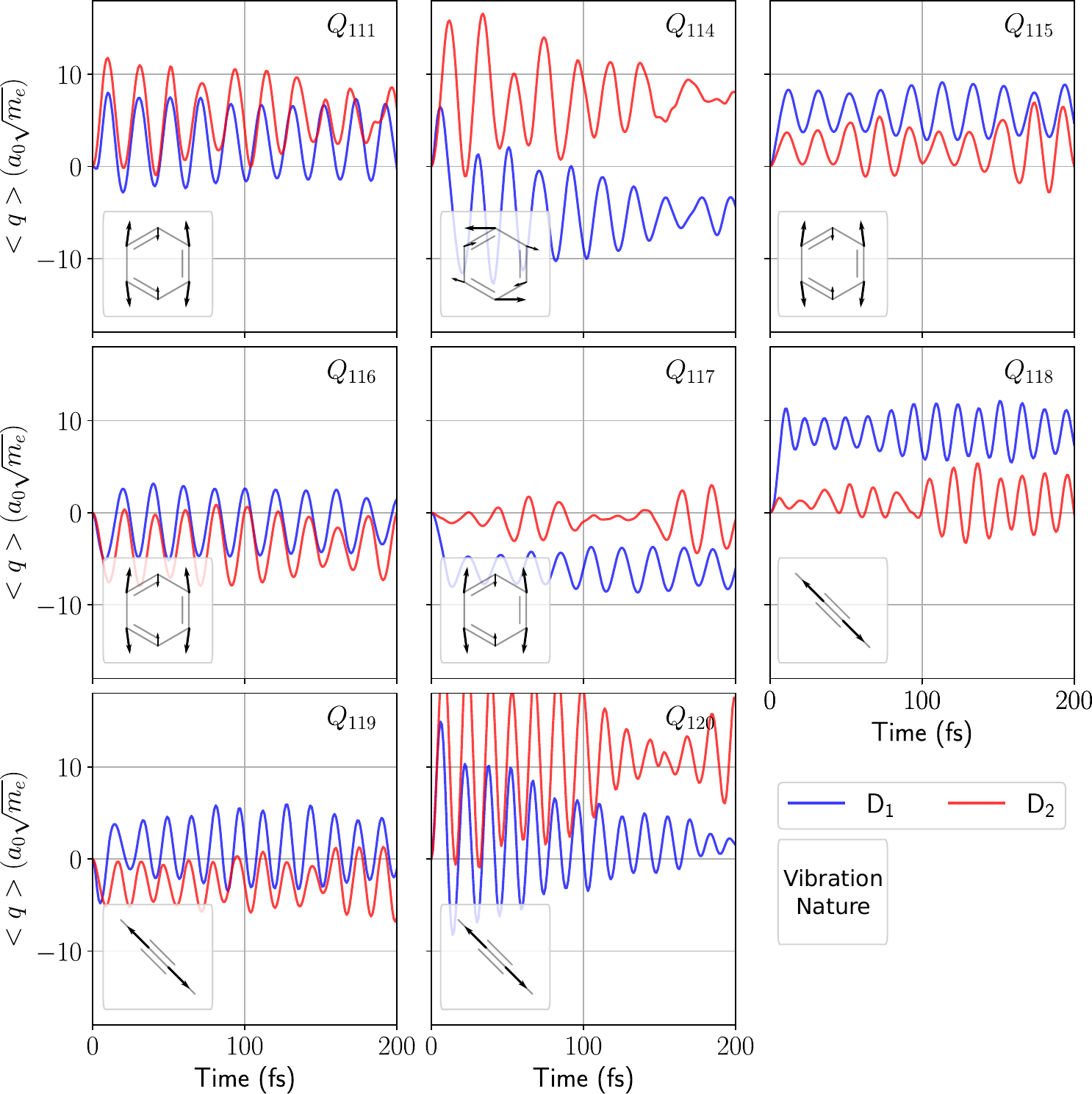}
  \includegraphics[width=0.48\linewidth]{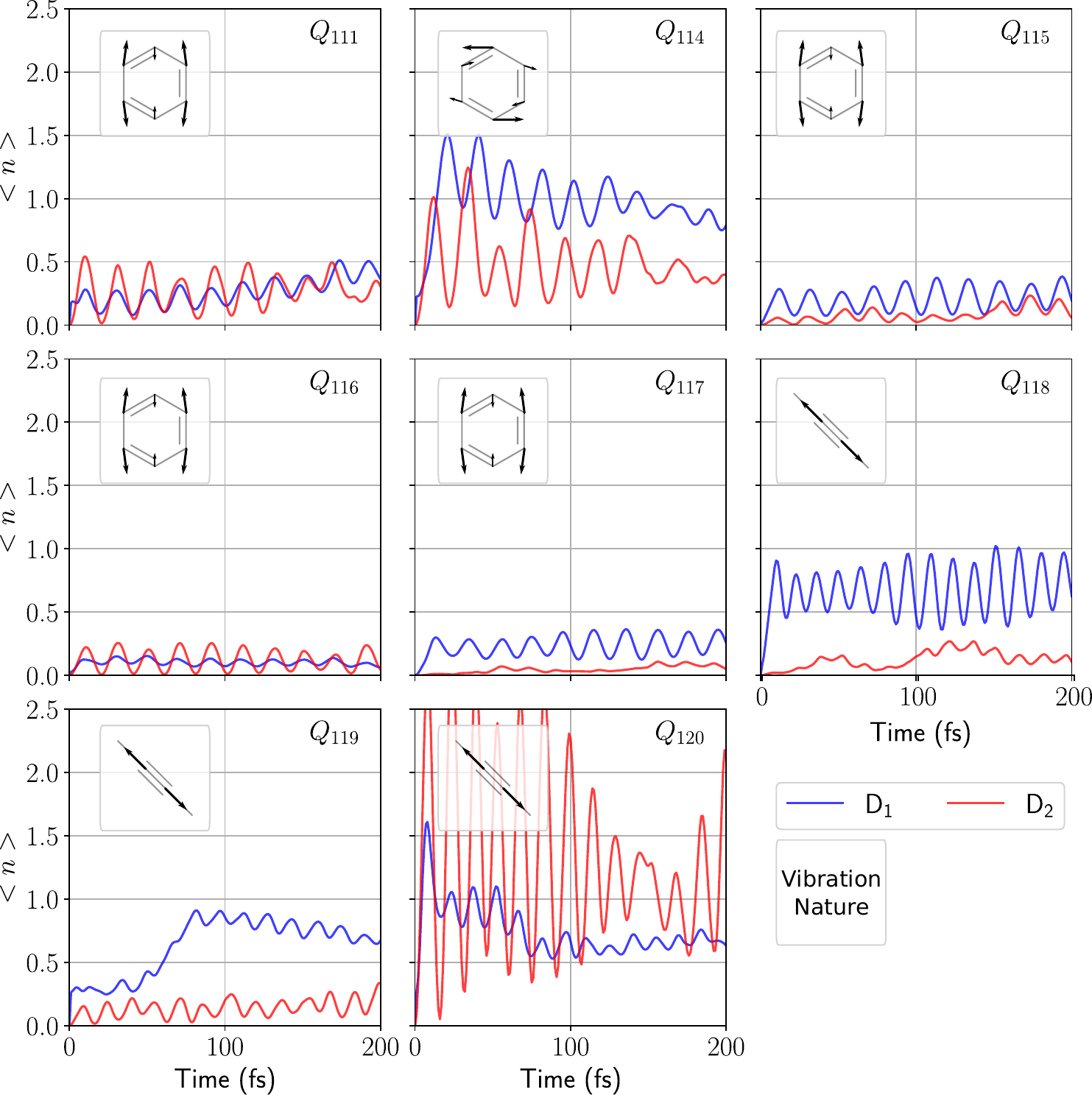}
  \caption{
    Time evolution of the expectation values of positions (left, mass-weighted atomic units), and vibrational excitation numbers (right) in diabatic states D\textsubscript{1} (blue line) and D\textsubscript{2} (red line) for an initial excitation on D\textsubscript{2}, for each normal mode.
    The nature of the vibration (quinoidal/acetylenic and stretching/rock-bending) is recalled for each normal mode; for the definition of the normal modes, see \cref{fig:quinace_modes_schematic}.
    The same quantities and their population-weighted variants are given for both sudden excitations in the SI (fig. SI5, SI6 and SI7).
    }
  \label{fig:exc_23_QNExp}
\end{figure*}

In our approach, the nuclear displacements are evaluated ultimately using nuclear wavepackets expanded in a basis of HO eigenfunctions.
The mean expectation value of the position operator for a given mode $\Braket{q_{i}}(t)$ is to be understood as the average center of the wavepacket. Such a quantity can be evaluated for each mode $i$ and also in each diabatic state $s$: $\Braket{q_{i}}_{s}(t)$ (\cref{fig:exc_23_QNExp}, left, after excitation of D\textsubscript{2}).

The latter expectation value is intensive and does not depend on the population of the considered state.
As such, it relates to the value of the displacement $q_{i}$ being considered as an observable at a given time, if it were measurable in each diabatic state $s$.

Another relevant quantity for characterizing the nuclear wavepacket is the vibrational excitation number $\Braket{n_i}_{s}(t)$ for each mode $i$ and each state $s$.
Again, we can look at these numbers for each diabatic state (\cref{fig:exc_23_QNExp}, right, after excitation of D\textsubscript{2}).

The initial excitation of diabatic state D\textsubscript{1} yields little significant variations in the positions of the wavepacket center.
More precisely, the variations are (quasi-)periodic with a typical oscillation time corresponding to that of each normal mode.
The same observation is made for vibrational numbers (fig. SI5 and SI6).

Furthermore, the dispersions (standard deviations) for both quantities (denoted $\Braket{\mathrm{d}q_i}_s$ and $\Braket{\mathrm{d}n_i}_s$) are almost constant, involving that the wavepacket mostly maintains its shape during the propagation.
This is to be expected, since the numerical experiment of exciting the D\textsubscript{1} state leads to a dynamics initially driven by the important gradients of D\textsubscript{1} ($\simeq$ S\textsubscript{1}), such that the wavepacket is instantaneously trapped in the lowest electronic state with negligible coupling and no re-crossing to the D\textsubscript{2} state.
The observed erratic variations of $\Braket{q_i}_s(t)$ and $\Braket{n_i}_s(t)$ in the D\textsubscript{2} diabatic state are not physically relevant because almost zero population is in the D\textsubscript{2} diabatic state (fig. SI5 and SI6; maximum population 0.05, at the initialization).

On the other hand, the initial excitation of D\textsubscript{2} yields much more interesting results for EET.
First, for most of the expectation values, we retrieve expected oscillations with two frequency-separated groups: quinoidal (111, 114, 115, 116, 117) and acetylenic (118, 119, 120) degrees of freedom.
During the first few \unit{\femto\second}, initial excitation on diabatic state D\textsubscript{2} involves strong displacements along the central anti-quinoidal mode 114 and the p2-acetylenic mode 120, which strongly participate in the gradient difference vector (\cref{tab:mode_contributions_model}).
Both modes also participate in the inter-state coupling vector such that, in conjunction with EET population transfer, their excitation is transferred into diabatic state D\textsubscript{1}.
The p3-acetylenic mode 118 increases concomitantly in the diabatic state D\textsubscript{1} apart from these oscillations.
To a lesser extent, displacements associated to the p3-quinoidal mode 117, also localized on the p3-branch, show a similar behavior.
This corresponds to the displacement of the wavepacket toward the equilibrium geometry of the first diabatic states once it is populated.
In other words, the time evolution of all $\Braket{q_{i}}_{s}$ for both states is consistent with the role of the corresponding modes within the plane spanned by the approximate branching-space vectors $(\mathbf{g}',\mathbf{h}')$.

The vibrational excitation numbers can be interpreted in the same way.
Initially, the p2-acetylenic mode 120 is highly excited (oscillating around $\Braket{n_{120}}=1$), along with the central-quinoidal mode 114.
This vibrational excitation is transferred within the first \qty{25}{\femto\second} toward p3-localized modes on diabatic state D\textsubscript{1} (acetylenic modes 118 and 119).

\begin{figure}
\includegraphics[width=0.85\linewidth]{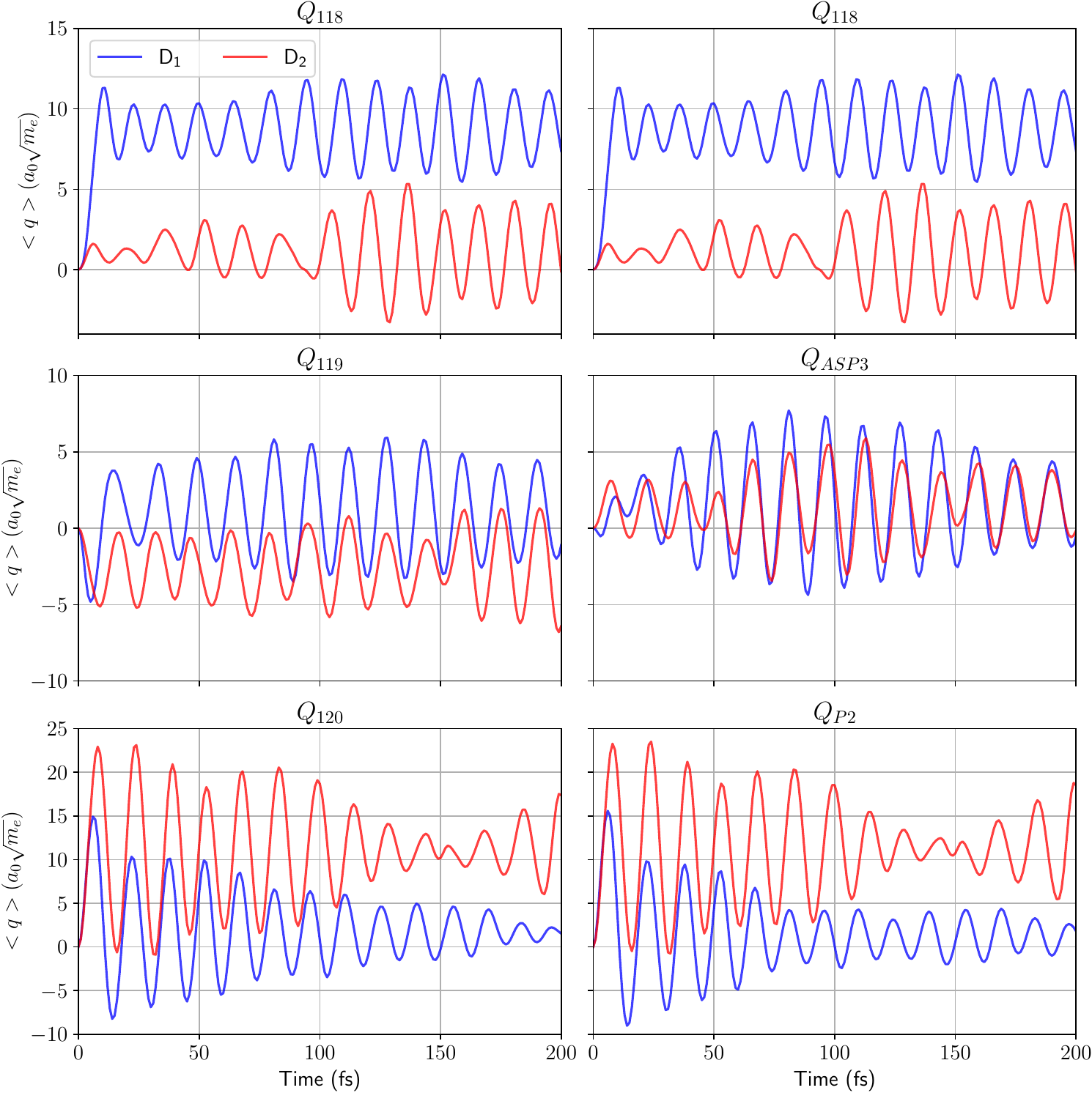}
\caption{
  Time evolution of the expectation values of position in diabatic states D\textsubscript{1} and D\textsubscript{2} for an initial excitation on D\textsubscript{2}, for normal modes (118, 119, 120) on the left and modes (118, $Q$\textsubscript{p2}, $Q$\textsubscript{ASp3}) on the right.
  }
\label{fig:fully_localized_acetylenic_modes}
\end{figure}

However, the study of the normal modes of vibration is not completely satisfactory for the study of the EET between localized excitons.
Indeed, the normal modes of vibration associated to elongations of the acetylenic bonds localized on the p3 branch are not completely separated from p2.
As shown in \cref{fig:quinace_modes_schematic}, mode 118 is fully localized on the p3 branch, but modes 119 and 120 are only partially localized on the p2 branch.
The pair of normal modes 119 and 120 can be further transformed into a pair of local modes (which are not anymore normal modes for m23) that have almost zero displacements on the p3 branch.
This re-localization of the normal modes gives fragment vibrations:
\begin{itemize}
\item a fully-localized elongation of the acetylenic bond on the p2 branch ($Q$\textsubscript{p2});
\item a fully-localized asynchronous elongation of the two acetylenic bonds on the p3 branch ($Q$\textsubscript{ASp3});
\end{itemize}
with the latter being the counterpart of the fully-localized synchronous elongation of the two acetylenic bonds on the p3 branch (mode 118).
Same as before, we can look at the center of the wavepackets for these modes and compare before and after localization, \cref{fig:fully_localized_acetylenic_modes}.
We notice a quantum-beat-like oscillation for $\Braket{q_{\text{ASp3}}}(t)$, but in low amplitudes, with a periodic envelop of about \qty{200}{\femto\second}.
$\Braket{q_{\text{p2}}}(t)$ resembles the evolution of 120 because the coefficient in the linear combination of 119 and 120 is much greater for 120.
We interpret this as 120 being mostly an acetylenic elongation on p2, contaminated with some asynchronous elongation on p3.
\subsection{Internal coordinates analysis of EET}
For better comparison with the previously published direct-dynamics simulations on systems such as m23 or related, we now illustrate the evolution of the internal coordinates during the EET process.

The most relevant internal coordinates for PPEs and analogous systems have been identified as the lengths of the acetylenic bonds.\cite{fernandez-alberti_nonadiabatic_2009,negrin-yuvero_vibrational_2023}
Such quantities are obtained using the displaced coordinates
\begin{subequations}
  \begin{equation}
\Delta \mathbf{R}^{(s)}(t)
=\sum_{i\text{, modes}}\frac{\Braket{q_i}_s(t)}{\sqrt{\mu_{i}}}\mathbf{L}_{\text{Cart,}i} \quad,
  \end{equation}
  \begin{equation}
\Delta \mathbf{R}(t)
=\sum_{i\text{, modes}}\frac{\Braket{q_i}(t)}{\sqrt{\mu_{i}}}\mathbf{L}_{\text{Cart,}i} \quad,
  \end{equation}
\end{subequations}
where $\mathbf{L}_{\text{cart,i}}$ and $\mu_{i}$ are the Cartesian displacements and reduced mass associated to the normal mode $i$.
The state-specific displaced coordinates $\Delta \mathbf{R}^{(s)}(t)$ and total displaced coordinates $\Delta\mathbf{R}(t)$ are calculated by adding the Cartesian displacements of the normal modes weighted by the position of center of the wavepacket.
Relevant distances are then easily computed knowing the atom numbering in the molecule.
We focus here on the lengths of the acetylenic bonds.

The equilibrium length of the acetylenic bond is \qty{1.21}{\angstrom} in the ground state of PPEs (with an alternated C-C$\equiv$C-C bonding pattern), and \qty{1.25}{\angstrom} in the first excited state (toward a cumulenic C=C=C=C bonding pattern).

\begin{figure}
\includegraphics[width=1.0\linewidth]{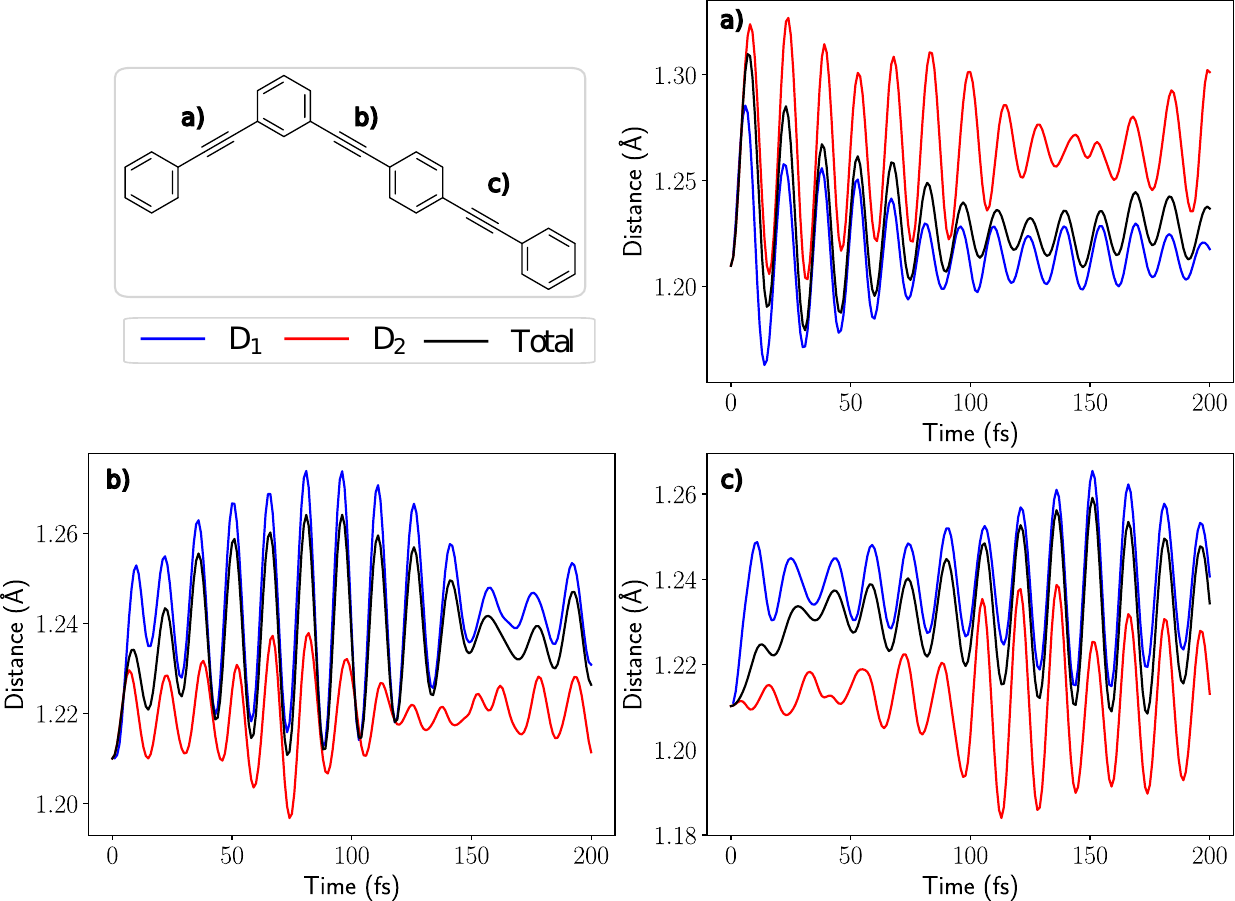}
\caption{Time evolution of the lengths of the acetylenic bonds of m23 in either of the two diabatic states D\textsubscript{1} or D\textsubscript{2} (blue and red lines), or in total (black line).}
\label{fig:acetylenic_bond_lengths}
\end{figure}

We see that during the early dynamics the p2 acetylenic bond is highly distorted (\cref{fig:acetylenic_bond_lengths}, \textbf{(a)}), elongated beyond the equilibrium geometry of MinS\textsubscript{1}, and oscillates back toward the equilibrium geometry of MinS\textsubscript{0}.
During the same time (the first \qty{25}{\femto\second}), the "internal" p3 acetylenic bond (closest to the central phenylene) is elongated but exhibits large-amplitude oscillations with a period of about \qty{150}{\femto\second} (\cref{fig:acetylenic_bond_lengths}, \textbf{(b)}).
The "external" (hence peripheral) p3 acetylenic bond is elongated in the same time scale (first \qty{25}{\femto\second}) but the high-amplitude oscillations begin only at \qty{50}{\femto\second}, with an analogous period of \qty{150}{\femto\second} (\cref{fig:acetylenic_bond_lengths}, \textbf{(c)}).

From a local, internal bonds, point of view, the excitation on the second excited state D\textsubscript{2} yields EET dynamics with two main features:
i) strong elongation of the p2 acetylenic bond; ii) moderate elongation of the two p3 acetylenic bonds with asynchronous oscillations between the two bond lengths.
These are consistent with previous studies of internal coordinates evolution during EET in m23.\cite{fernandez-alberti_nonadiabatic_2009}

In the  latter work, the authors used mixed quantum-classical dynamics with propagation of multiple trajectories and surface-hopping.
The trajectories are separated into two sets: effective (non back-hopping after) or non-effective (existing back-hopping) hop from S\textsubscript{2} to S\textsubscript{1}.
Our results are consistent with the set of effective-hop trajectories.

Note that the large-amplitude oscillations for the p3 acetylenic bonds are maintained for about \qty{600}{\femto\second} in our simulations, with four complete beats (validating a period of about \qty{150}{\femto\second}) (see late dynamics in \cref{sec:ivr}).

\subsection{Internal Vibronic Redistribution}
\label{sec:ivr}
\subsubsection*{Electronic and Vibrational partitions}
We also analyze the EET process in m23 through the lens of the internal redistribution of energy.
First, let us discuss the overall partition of energy within the system.

The total diabatic energy (which is not the total energy in our definition) is the sum of the expectation values for all diagonal Hamiltonian operators in the two-state formalism:
\begin{equation}
  E_{\text{diabatic}}(t)=E_{\text{electronic}}(t)+E_{\text{vibrational}}(t) \quad,
\end{equation}
where neither of the three quantities are conserved.
The total energy of the system is, however, duly conserved and takes into account the contribution of the off-diagonal diabatic terms, $E_{\text{off-diagonal}}(t)$.

The electronic and vibrational energy are further decomposed as 
\begin{subequations}
  \begin{equation}
    \begin{aligned}
    E_{\text{electronic}}(t)=
    &P_{1}(t)E^{(1)}(t=0,\mathbf{Q}=0)\\
    +
    &P_{2}(t)E^{(2)}(t=0,\mathbf{Q}=0)
    \end{aligned} \quad,
  \end{equation}
  \begin{equation}
    \begin{aligned}
    E_{\text{vibrational}}(t)=
      \sum_{s\text{, states}}
      P_{s}(t)
    \sum_{i\text{, modes}}
    &\Braket{T_{\text{nu,}i}}_{s}\\
    +
    &\kappa_{i}^{(s)}\Braket{q_{i}}_{s}(t)\\
    +
    & \frac 1 2 k_{i}^{(s)}\Braket{q_{i}^{2}}_{s}(t)
    \end{aligned} \quad.
    \label{eq:vibrational_decomposition}
  \end{equation}
\end{subequations}
With these definitions, the electronic energy is the reservoir of energy in the constant part of the diabatic potentials at the origin ($\mathbf{Q}=0$) and the vibrational energy is the energy resulting from the nuclear displacements from the origin.

\begin{figure*}
  \includegraphics[width=1.0\linewidth]{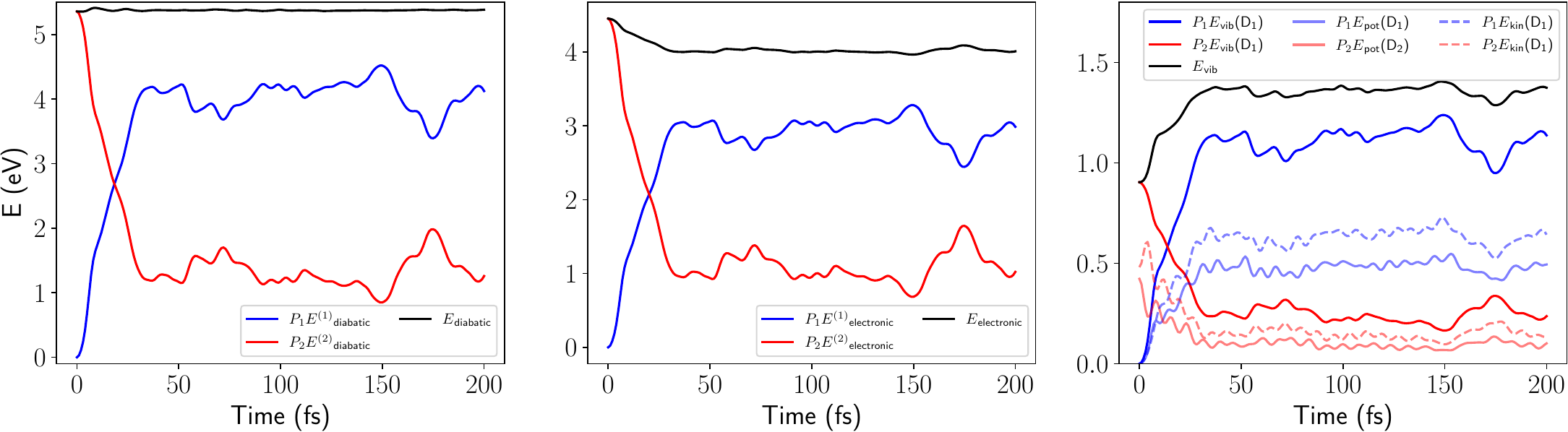}
  \caption{
    Time evolution of the diabatic (diagonal) energy and its decomposition.
    Contributions from diabatic states D\textsubscript{1} and D\textsubscript{2} are blue and red lines, respectively.
    Left: contributions to the total diabatic energy (black line).
    Center: contributions to the total electronic energy (black line).
    Right: contributions to the total vibrational energy (black line) from potential energy (transparent lines) and kinetic energy (dashed lines).
  }
  \label{fig:energy_decomposition_DiabaticElVib}
\end{figure*}

We illustrate the energy decomposition in \cref{fig:energy_decomposition_DiabaticElVib}.
From left to right, we show the diabatic, electronic, and vibrational energies in the system.
Again, the total diabatic energy is not strictly conserved.
The conservation of the total energy is illustrated in the SI with the total diabatic energy and the contribution of the off-diagonal terms (fig. SI8).

The total electronic energy decreases from \qty{4.45}{\electronvolt} (initial vertical transition energy) to approximately \qty{3.9}{\electronvolt} \cref{fig:energy_decomposition_DiabaticElVib} (center panel, black line). 
The excess energy is transferred to the vibrational wavepacket as a vibrational energy \cref{fig:energy_decomposition_DiabaticElVib} (right panel, black line).

\begin{figure}
  \includegraphics[width=0.8\linewidth]{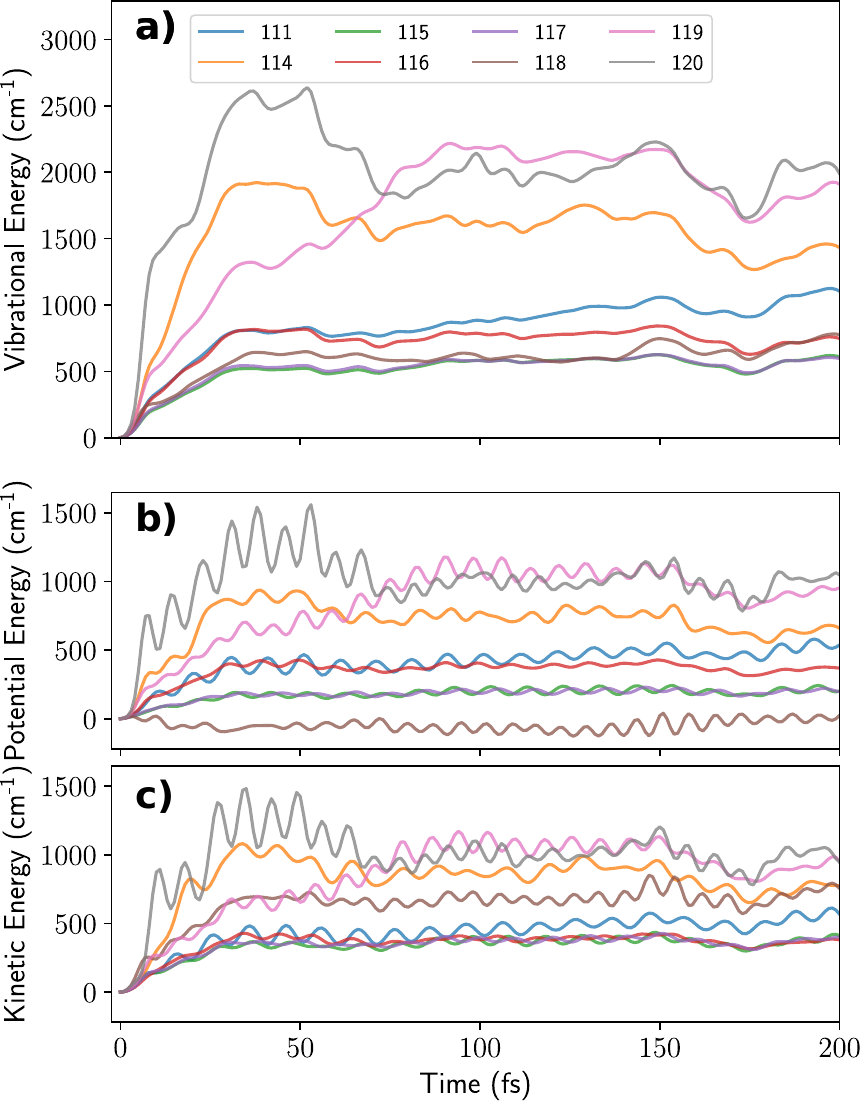}
  \caption{
    Time evolution of the vibrational, potential and kinetic energy mode per mode for contributions from diabatic states D\textsubscript{1}.
    The equivalent contributions from diabatic state D\textsubscript{2} are given in the SI (fig. SI9).
  }
  \label{fig:energy_decomposition_VibPerMode}
\end{figure}

The decomposition of the vibrational energy in \cref{eq:vibrational_decomposition} allows for a discussion of the energy per state and per mode.
We can thus estimate the distribution of the excess vibrational energy in the different normal modes, comparing for each mode the kinetic energy, potential energy, and vibrational energy.
The contributions of such quantities from the diabatic state D\textsubscript{1} are given in \cref{fig:energy_decomposition_VibPerMode}.

At $t=\qty{0}{\femto\second}$, all the vibrational energy is in state D\textsubscript{2}, and is equal to the zero-point energy (ZPE) associated to the vibrational ground state of the ground electronic state of the molecule.
The initial vibrational energy is thus simply around \qty{820}{\per\centi\meter} and \qty{1150}{\per\centi\meter} for quinoidal and acetylenic modes, respectively (see $t=\qty{0}{\femto\second}$ in \cref{fig:energy_Vib_1000fs} or fig. SI9).

In the first \qty{25}{\femto\second}, the vibrational energy in state D\textsubscript{2} is transferred to D\textsubscript{1}.
The vibrational energies in D\textsubscript{2} of each mode are given in the SI (SI fig 8) and quickly tend to an almost-zero residual energy, consistently with the D\textsubscript{2} vibrational energy, see \cref{fig:energy_decomposition_DiabaticElVib}, right.
Only the p2-acetylenic mode 120 exhibits significant potential energy and kinetic energy variations.
On the other hand, the vibrational distribution in D\textsubscript{1} is no longer like a ZPE.

In particular, the excess energy from the electronic reservoir is split into two groups, with spectator modes (115, 116, 117 and 118) and active (excited) modes (111, 114, 119 and 120) (panel \textbf{(a)} in \cref{fig:energy_decomposition_VibPerMode}).

This energy distribution relates to the vibrational excitation numbers computed in \cref{fig:exc_23_QNExp}.
The first set of modes (p2- and p3- quinoidal-modes, and the synchronous p3-acetylenic mode) stabilizes below \qty{800}{\per\centi\meter}, corresponding to almost zero vibrational excitation.
Mode 120 is highly excited in the initial state D\textsubscript{2},  and its excitation remains in D\textsubscript{1} (most energetic mode), which denotes its importance in the reduced model for energy redistribution.

Now, the excitation remaining in D\textsubscript{1} for mode 120 explains the oscillations in the diabatic populations and low diabatic quantum yield.
The displacements induced by the significant vibrations of mode 120 bring the system closer to the intersection seam and prevent a complete relaxation toward the D\textsubscript{1} state.

The next two most energetic modes are the p3-asynchronous-acetylenic mode 119 and the central anti-quinoidal mode 114.
Their energy stabilizes later, around \qty{100}{\femto\second}, in D\textsubscript{1}.
Finally, the vibrational energy of the central quinoidal mode 111 increases as part of the spectating modes, but stabilizes as an excited mode.
It is clear from \cref{tab:mode_contributions_model} that modes contributing strongly to the inter-state coupling vector keep a significant vibrational excitation after the electronic transfer.
On the contrary, the spectating modes from the vibrational excitation perspective are only involved in the diagonal gradient vectors (both energy difference and gradient average vectors).

From this internal vibrational redistribution of the excess electronic excitation, we identify a segregation between active and spectator modes within our model.
This is consistent with mixed quantum-classical trajectory-based dynamics calculations in analogous systems 
\cite{alfonso-hernandez_vibrational_2020}, for which two groups of modes are identified: active modes, which transiently store the excess energy during EET, and spectator modes, which provide a bath of lowly excited modes and contribute to dissipate the excess energy in the late dynamics.
However, with the present model, the expected thermal equilibration of the excess energy among all the normal modes is not reached at the end of the simulations, as will be pointed out in the next section.
\subsubsection*{A remark on Late Dynamics}
\begin{figure}
  \includegraphics[width=0.8\linewidth]{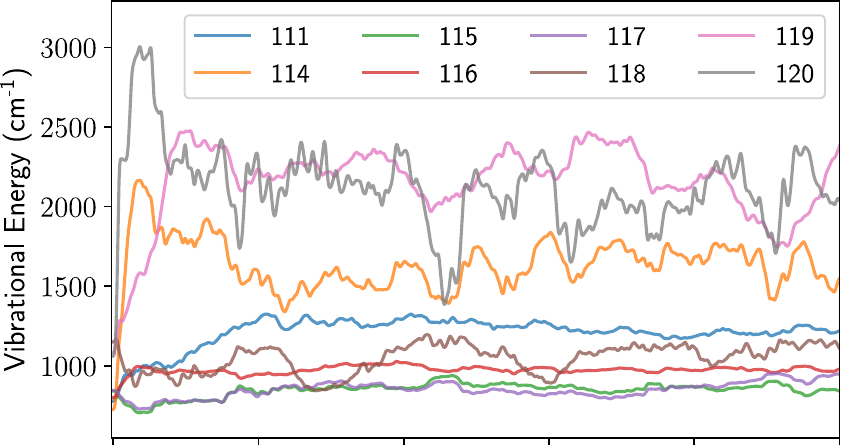}
  \caption{
    Time evolution of the vibrational energy per mode from both diabatic states D\textsubscript{1} and D\textsubscript{2}.
  }
  \label{fig:energy_Vib_1000fs}
\end{figure}

The wavepackets were propagated up to \qty{1000}{\femto\second} to evaluate possible thermalization-like effects in late dynamics.
Let us remind that the effective population transfer takes place very early, in the first \qty{25}{\femto\second}.
We provide the time evolution of the vibrational energy at long times in \cref{fig:energy_Vib_1000fs}.

We can still observe two groups of \textit{energy-spectator} modes (now 111, 115, 116, 117 and 118) and \textit{active} modes (114, 119 and 120).

Because of the low-dimensionality of the model and of the absence of interaction with the environment or low-frequency modes, we cannot really reach complete thermal equilibrium.
Tuning-only modes appear to reach equilibrium with equipartition of energy within the group (the four lowest-energy modes).
The modes strongly involved in the inter-state coupling with the electronic reservoir are still out-of-equilibrium at long times.

\begin{figure}
\includegraphics[width=1.0\linewidth]{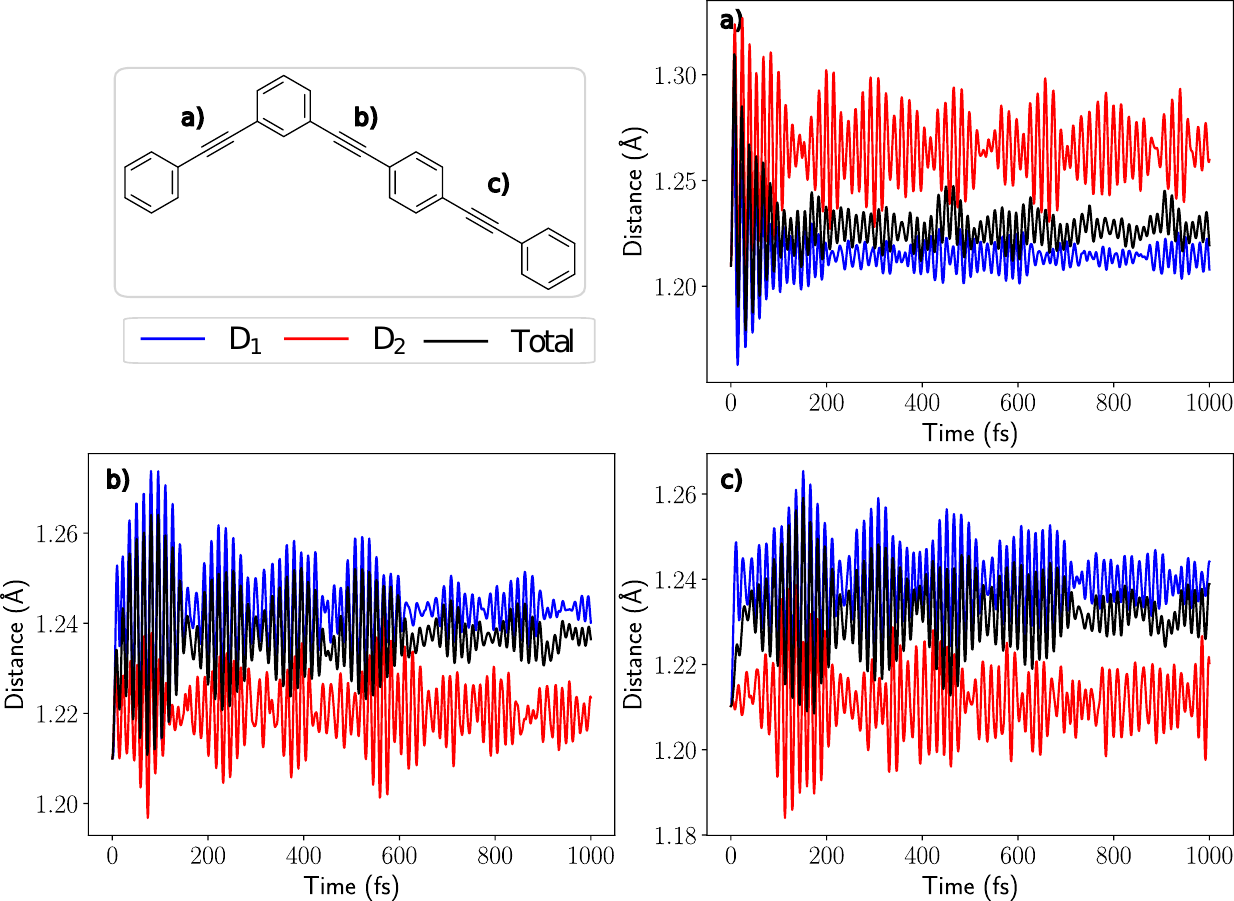}
\caption{
  Time evolution of the lengths of acetylenic bonds of m23 in either of the two diabatic states D\textsubscript{1} or D\textsubscript{2} (blue and red lines) or in total (black line).
  }
\label{fig:acetylenic_bond_lengths_late}
\end{figure}

In addition, the asynchronous oscillations of the acetylenic bonds lengths in the p3-branch are damped to some extent after four periods (\cref{fig:acetylenic_bond_lengths_late}, \textbf{(b, c)}).

\section{Conclusions and Outlook}
\label{sec:conclusion}
In the present work, we addressed with quantitative quantum computations the stationary and time-resolved properties of the first asymmetrically \textit{meta}-substituted PPE oligomer, named m23.

The EET process from $\text{S}_{2}$ to $\text{S}_{1}$ (LE electronic states localized on the p2 and p3 pseudo-fragments, respectively) was characterized extensively from a vibronic perspective involving nonadiabatic passage through conical intersections.

To this end, we derived a simple -- yet not based on symmetry -- diabatic LVC Hamiltonian model for a dimensionally-reduced (8 out of 138 vibrations; 1+2 coupled electronic states) description of the system.

Such a parameterized Hamiltonian was used to run quantum dynamics simulations with MCTDH wavepacket propagations. It was proved to give a satisfactory description of the initially populated Franck-Condon region, together with a decent estimation of the strength of the inter-state coupling around the conical intersection seam, and thus able to describe the overall EET process realistically.

With this, EET was estimated to occur very efficiently within about \qty{25}{\femto\second} after initial excitation to the highest-lying excited state of the model.
Population inversion was shown to occur in conjunction with significant nuclear displacements along the most coupling normal modes (on the p2 pseudo-fragment and central-phenylene modes).

After inversion and transfer, we showed that the system significantly shifts towards -- and oscillates around -- the equilibrium geometry of the lower-lying excited state (p3 pseudo-fragment). 
From a chemical point of view, vibrations of the p2 pseudo-fragment are thus transferred very fast but asynchronously to those of p3, concomitantly with the electronic population transfer. 
This can be interpreted through the lens of a simple decomposition of the total energy, providing a measure of the extent of internal vibrational redistribution (IVR). 

Accordingly, the quantum coherence between the two electronic states stays significant during the nonadiabatic transfer and correlated with the time evolution of the population difference for about \qty{40}{\femto\second}. 
It then decays when the system starts oscillating stochastically in the lowest-lying excited state, around its equilibrium geometry.

From a methodological point of view, our study based on wavepacket-based quantum dynamics and \textit{ab initio} data confirms nicely most of what has been inferred so far on such systems from previous studies in the literature based on mixed quantum-classical trajectory-based dynamics and semiempirical data (as regards the most active modes, as well as the rate and the yield of EET). 
This is consistent with an almost-ballistic crossing of the conical-intersection seam (sand-in-the funnel model). 
However, our observation of an -- in fact -- double crossing suggests some subtle effects that may require a full quantum treatment if one wants to address more specifically the detailed time evolution of electronic coherence here, which is induced nonadiabatically, and seems to persist for about a few tens of femtoseconds.

The present study could be improved on several fronts for further investigations.
The first limit is the assumption of a \emph{sudden excitation} picture for the initial states of the dynamics calculations.
A more rigorous treatment of the excitation from the electronic ground state, with the implementation of an explicit light-matter interaction (external electric field and molecular dipole).
This would allow a step toward realistic simulations of experiments, and could be compared to recent extensions of mixed quantum-classical trajectory-based molecular dynamics in the context of non-linear spectroscopy.\cite{hu_spectral_2021}

The second limit concerns the dimension reduction.
The environment (low-frequency modes, spectator modes, and possibly solvent modes) could be taken into account within a system-bath approximation.\cite{montorsi_spectroscopy_2023}
Further studies including a higher-dimensional model of the system may, indeed, provide a more realistic dynamical description of IVR in a system such as m23.

From a general perspective, we may expect electronic coherence to decay more strongly over longer times when adding extra degrees of freedom (and subsequent dissipation), but probably not so much faster during the ultrafast early dynamics (about a few tens of femtoseconds), which stays dominated by the few active vibrational modes considered in our model. 
The role of electronic coherence as a possible fingerprint for selecting EET candidates within high-dimensional systems remains to be further investigated.

\section*{Supplementary Material}
Supplementary material is available online for this work and upon request. 
It includes a complete set of ready-to-use parameters and operators files for the \texttt{Quantics} package.
The Cartesian coordinates in Angstrom for the minimum of the electronic ground state and for the minimum energy conical intersection are given.
Complementary figures for the expectation values of position and vibrational excitation operators for the photo-excitation toward the first diabatic state are also given.

\section*{Data Availability Statement}
The data that support the findings of this study are available from the corresponding author upon reasonable request.

\section*{Acknowledgments}
J. G. acknowledges the French MESR (Ministère de l’Enseignement Supérieur, de la Recherche) and the ENS (École Normale Supérieure) of Lyon for funding his PhD grant, hosted at the University of Montpellier.

\end{document}


\title[SI: Nonadiabatic EET through dendrimer building blocks]{Supplementary Information to:\\ Excitation-Energy Transfer and Vibronic Relaxation through Basic Light-Harvesting Dendrimer Building Blocks: a Nonadiabatic Perspective}


\author{Joachim Galiana}
\author{Benjamin Lasorne}%
\affiliation{ 
ICGM, Univ Montpellier, CNRS, ENSCM, Montpellier, France
}
\email{benjamin.lasorne@umontpellier.fr}
%


\date{\today}
\maketitle 

\section{Nuclear displacements along normal modes and branching-space vectors}
\begin{figure}[H]
\centering
\includegraphics[width=1\linewidth]{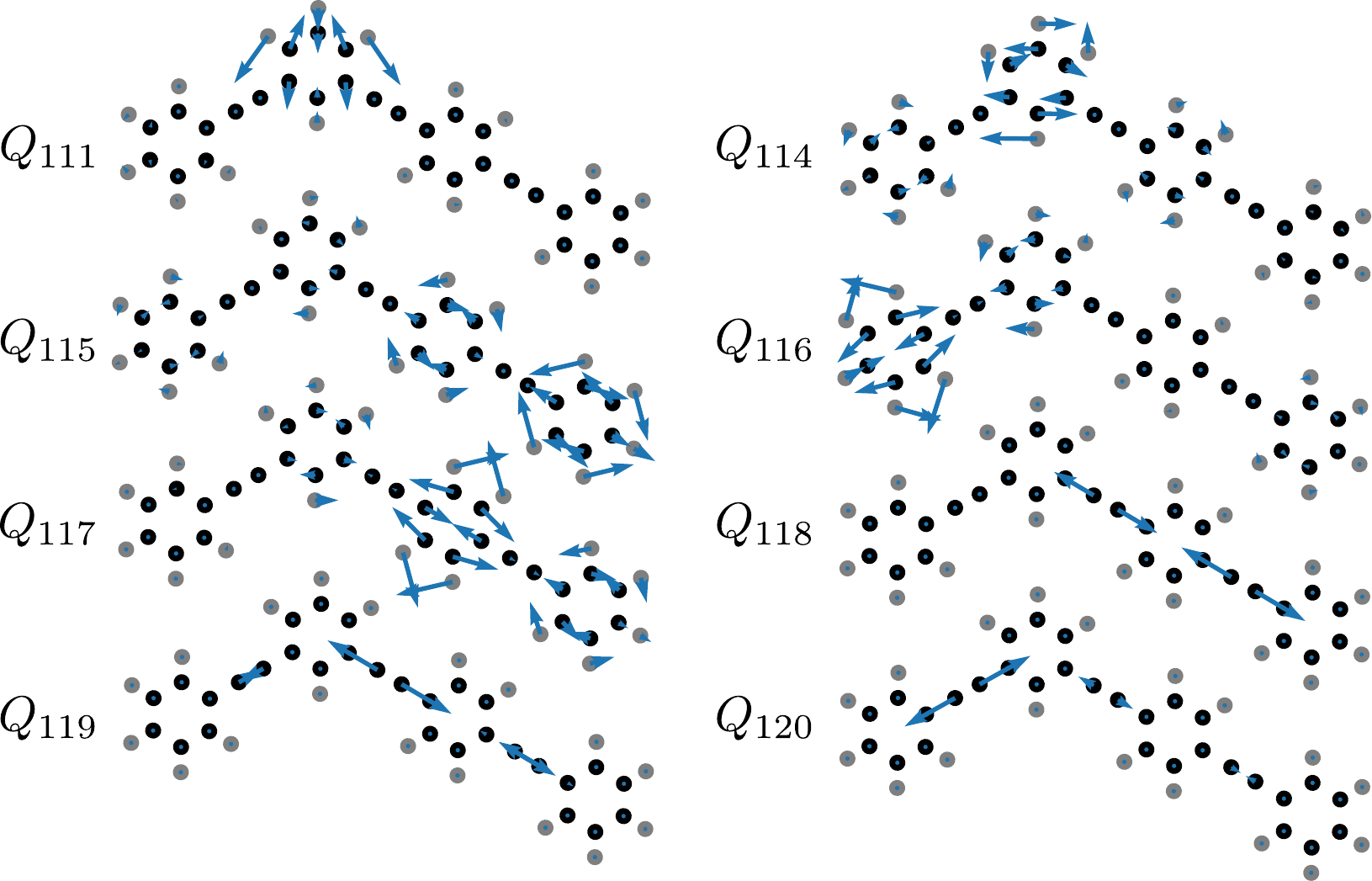}
\caption{
Cartesian displacements associated to the selected normal modes of vibration at the equilibrium geometry of m23 in its electronic ground state.
}
\label{fig:m23_QuinAce_modes_quivers}
\end{figure}
\begin{figure}[H]
\centering
\includegraphics[width=0.5\linewidth]{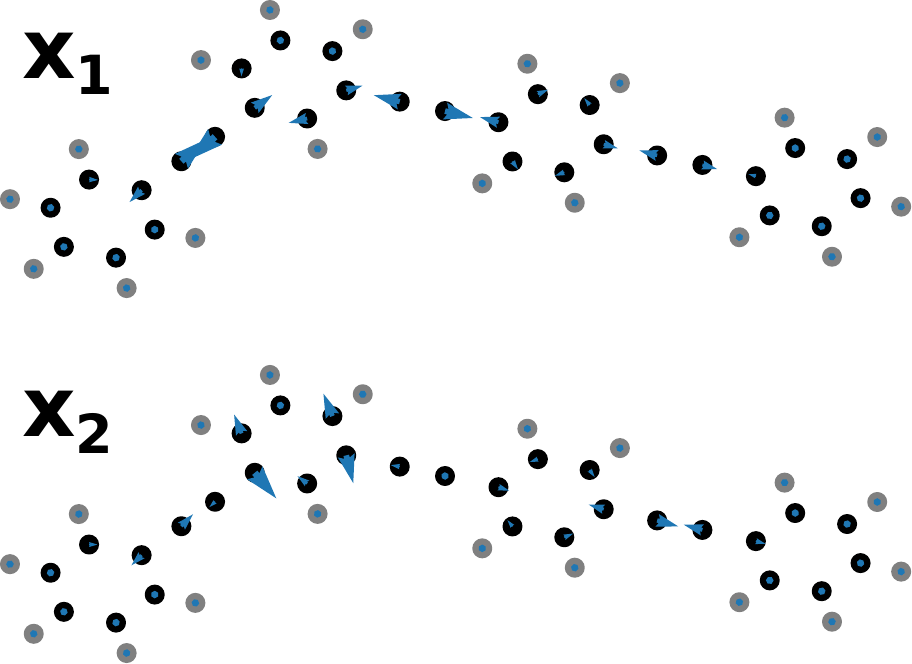}
\caption{
Cartesian displacements associated to the approximate branching-space vectors $\boldsymbol{x}_1\simeq \boldsymbol{g}$ and $\boldsymbol{x}_2\simeq\boldsymbol{h}$ at the MECI geometry.
}
\label{fig:hg_vectors_noRotation_quiver}
\end{figure}

\section{Linear Vibronic Coupling Model}
\begin{table}[H]
\centering
  \caption{
    LVC parameters obtained upon fitting \textit{ab initio} calculations.
    The parameters can be used unaltered in a \texttt{Quantics} operator file only setting reduced mass to 1 for each of the nuclear degrees of freedom.
    The diabatic energies at the reference point are $E^{(1)}=\qty{3.8766}{\electronvolt}$ and $E^{(2)}=\qty{4.4520}{\electronvolt}$.
  }
  \label{tab:mass_weighted_parameters}
  \begin{tabular}{lccrrr}
    \toprule
    Mode & $k_{i}^{(1)}$ & $k_{i}^{(2)}$ & $\kappa_{i}^{(1)}$ & $\kappa_{i}^{(2)}$ & $h'_i$ \\
    \midrule
    111 & 0.000056 & 0.000050 & -0.000143 & -0.000304 &  0.000096 \\
    114 & 0.000058 & 0.000043 &  0.000304 & -0.000313 & -0.000113 \\
    115 & 0.000056 & 0.000059 & -0.000337 & -0.000106 & -0.000034 \\
    116 & 0.000059 & 0.000053 &  0.000059 &  0.000192 &  0.000038 \\
    117 & 0.000056 & 0.000059 &  0.000345 &  0.000032 &  0.000010 \\
    118 & 0.000114 & 0.000110 & -0.001004 & -0.000108 &  0.000068 \\
    119 & 0.000092 & 0.000097 & -0.000107 &  0.000230 &  0.000141 \\
    120 & 0.000114 & 0.000092 & -0.000184 & -0.001025 & -0.000135 \\
    \bottomrule
  \end{tabular}
\end{table}

\begin{figure}[H]
\centering
\includegraphics[width=0.75\linewidth]{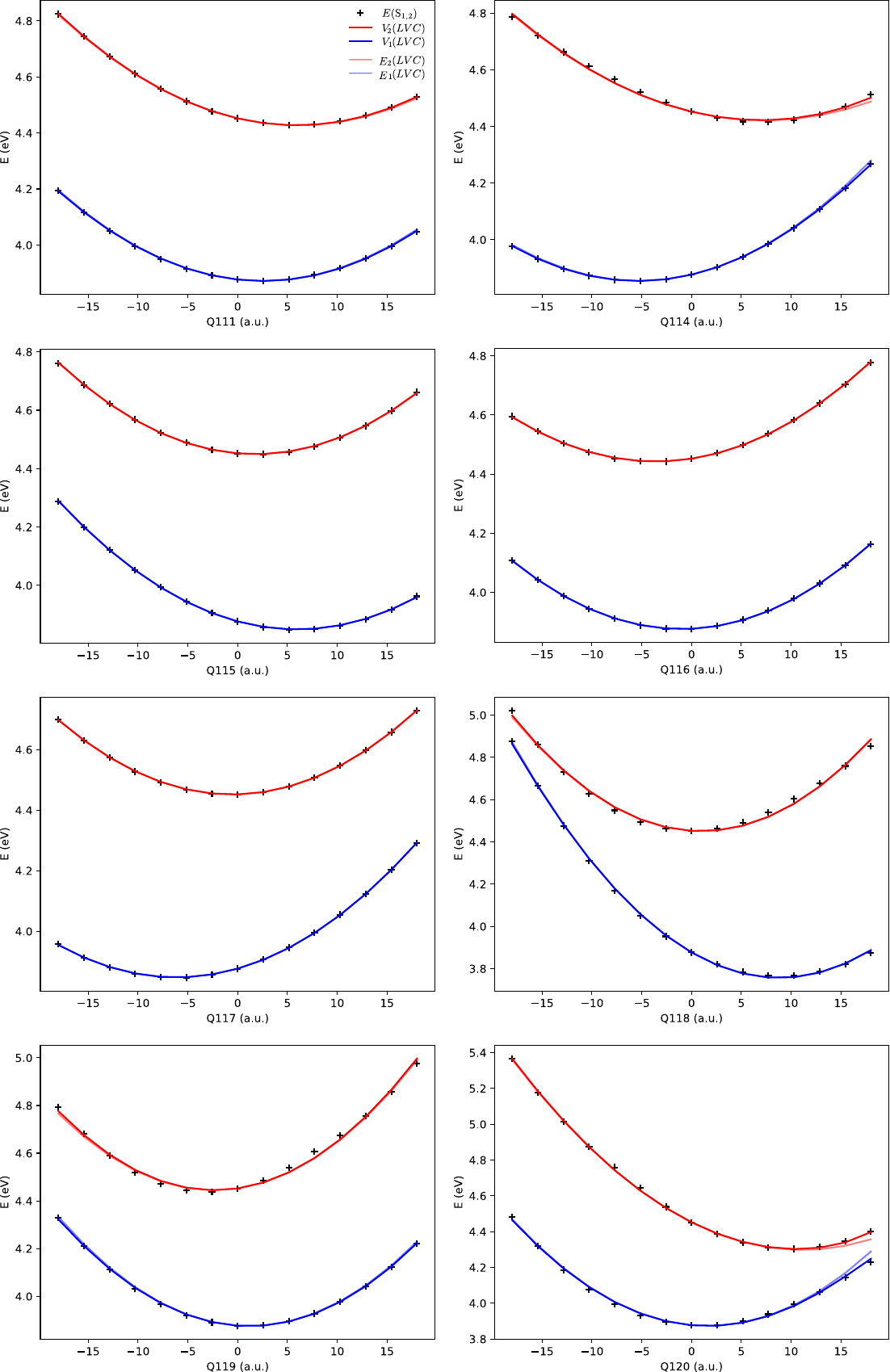}
\caption{
Rigid scans from the Franck-Condon geometry along normal mode displacements.
Black dots are the \textit{ab initio} energies for the first two adiabatic electronic excited states.
Plain lines are the first two eigenvalues of the LVC Hamiltonian model.
Transparent lines are the first two diabatic potentials of the LVC Hamiltonian model.}
\end{figure}

\begin{table}[H]
\centering
    \caption{Positions for the selected modes (with respect to the MinS\textsubscript{0} geometry) of the optimized critical points in the full system and in the dimensionally reduced model (in mass-weighted atomic units $a_0\sqrt{m_e}$).}
    \label{tab:LVC_critical_points_position}   \begin{tabular}{c|rrr|rrr}
    \toprule
         Mode &  Full-D MinS\textsubscript{1} & Full-D MinS\textsubscript{2} & Full-D MECI &  8-D MinS\textsubscript{1} & 8-D MinS\textsubscript{2} & 8-D MECI \\
         \midrule
         111  &   2.28 &    5.57 &  15.61 &  2.58 &  5.95 & 16.74 \\
         114  &  -5.17 &    4.57 &   3.79 & -5.43 &  7.19 &  4.51 \\
         115  &   5.76 &    2.19 &  -1.49 &  5.94 &  1.79 & -4.89 \\
         116  &  -1.10 &   -2.72 &  -0.40 & -0.97 & -3.57 & -0.99 \\
         117  &  -6.00 &   -1.15 &   1.82 & -6.14 & -0.54 &  4.88 \\
         118  &   8.07 &    1.00 &  -0.97 &  8.83 &  1.01 & -0.41 \\
         119  &   0.85 &   -2.36 &  -0.75 &  1.27 & -2.24 &  3.08 \\
         120  &   0.82 &    8.02 &   7.32 &  1.55 &  9.50 &  12.8 \\
    \bottomrule
    \end{tabular}
\end{table}

\begin{figure}[H]
  \includegraphics[width=0.75\linewidth]{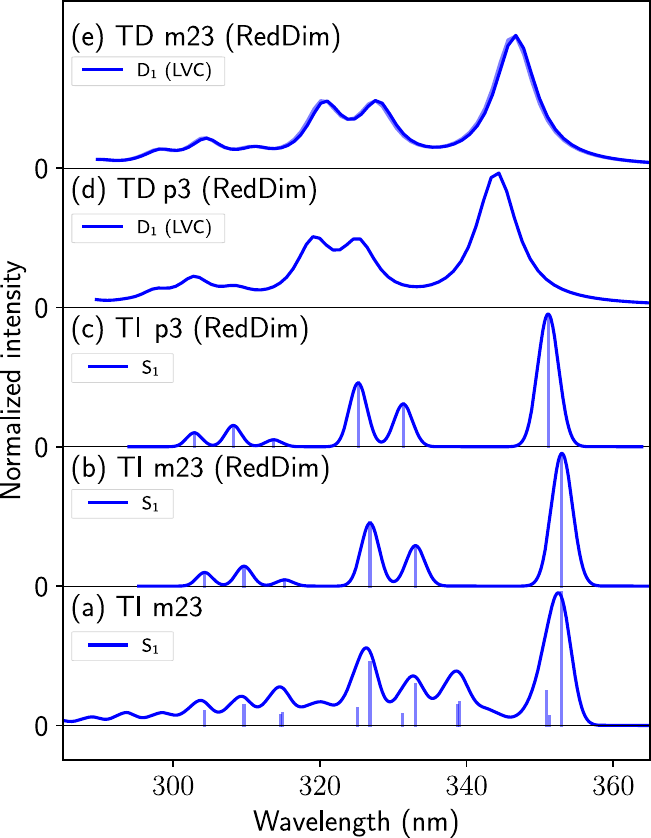}
  \caption{
    Calculated UV-visible spectra using a time-independent (TI) method giving Franck-Condon factors within the harmonic and Born-Oppenheimer approximations (a, b, c), and calculated power spectra using a time-dependent (TD) method beyond the Born-Oppenheimer approximation (d, e).
    \textbf{(a)} Full dimensional TI spectra between S\textsubscript{0} and S\textsubscript{1} of m23.
    \textbf{(b)} Reduced dimensional TI spectra between S\textsubscript{0} and S\textsubscript{1} of m23, selecting only transitions involving modes of the 8-D model.
    \textbf{(c)} Reduced dimensional TI spectra between S\textsubscript{0} and S\textsubscript{1} in p3 fragment.
    \textbf{(d)} Reduced dimensional TD power spectra between S\textsubscript{0} and D\textsubscript{1} in p3 fragment.
    \textbf{(e)} Reduced dimensional TD spectra between S\textsubscript{0} and D\textsubscript{1} of m23.
    The RedDim TD spectra \textbf{(d)} and \textbf{(e)} are shifted accordingly to the change in ZPE difference between the full system (138 modes) and the reduced model (8 modes).
  }
  \label{fig:all_absorption_spectra}
\end{figure}

\section{Quantum Dynamics Results}
\begin{figure}[H]
\centering
  \includegraphics[width=0.48\linewidth]{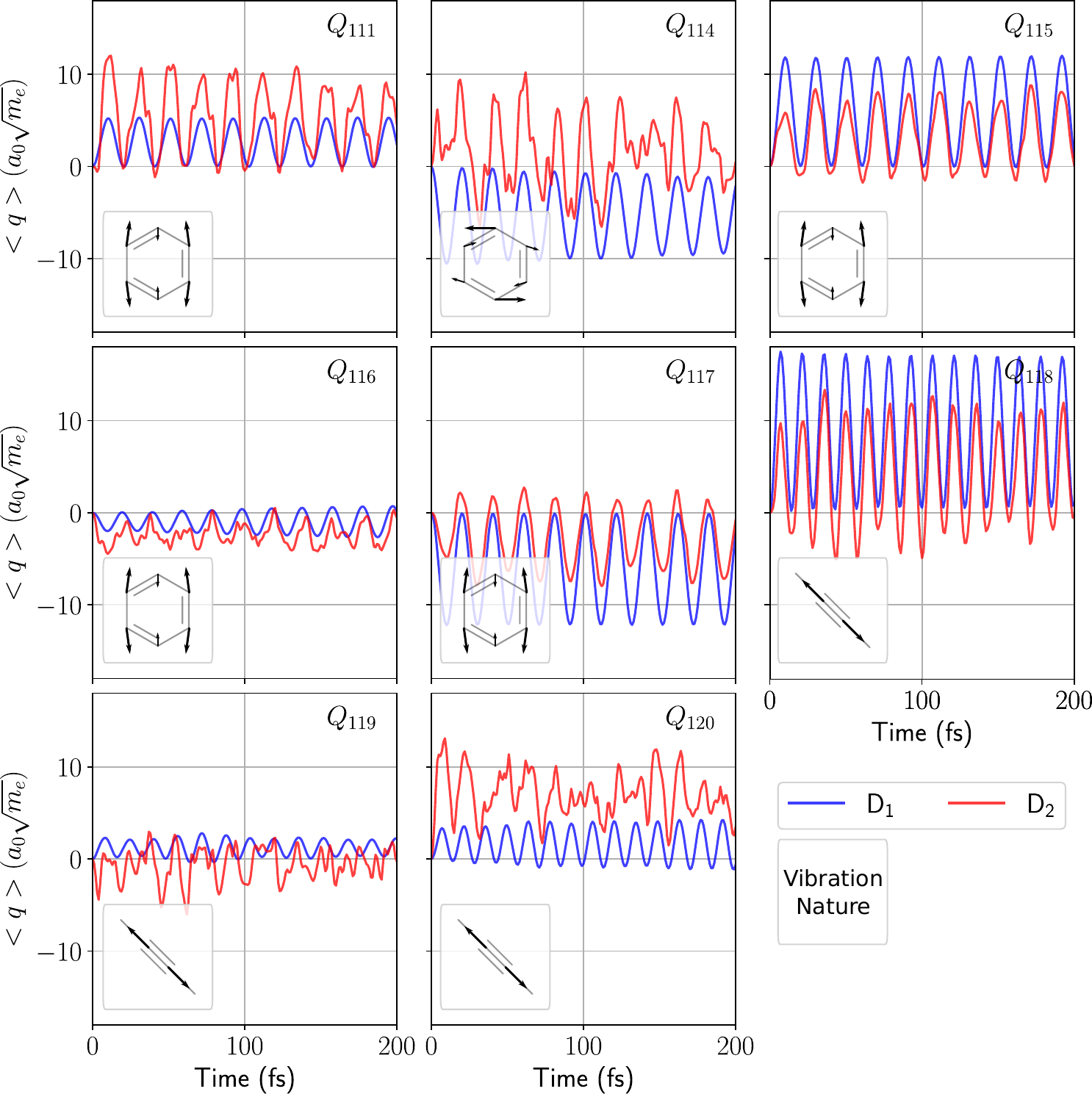}
  \includegraphics[width=0.48\linewidth]{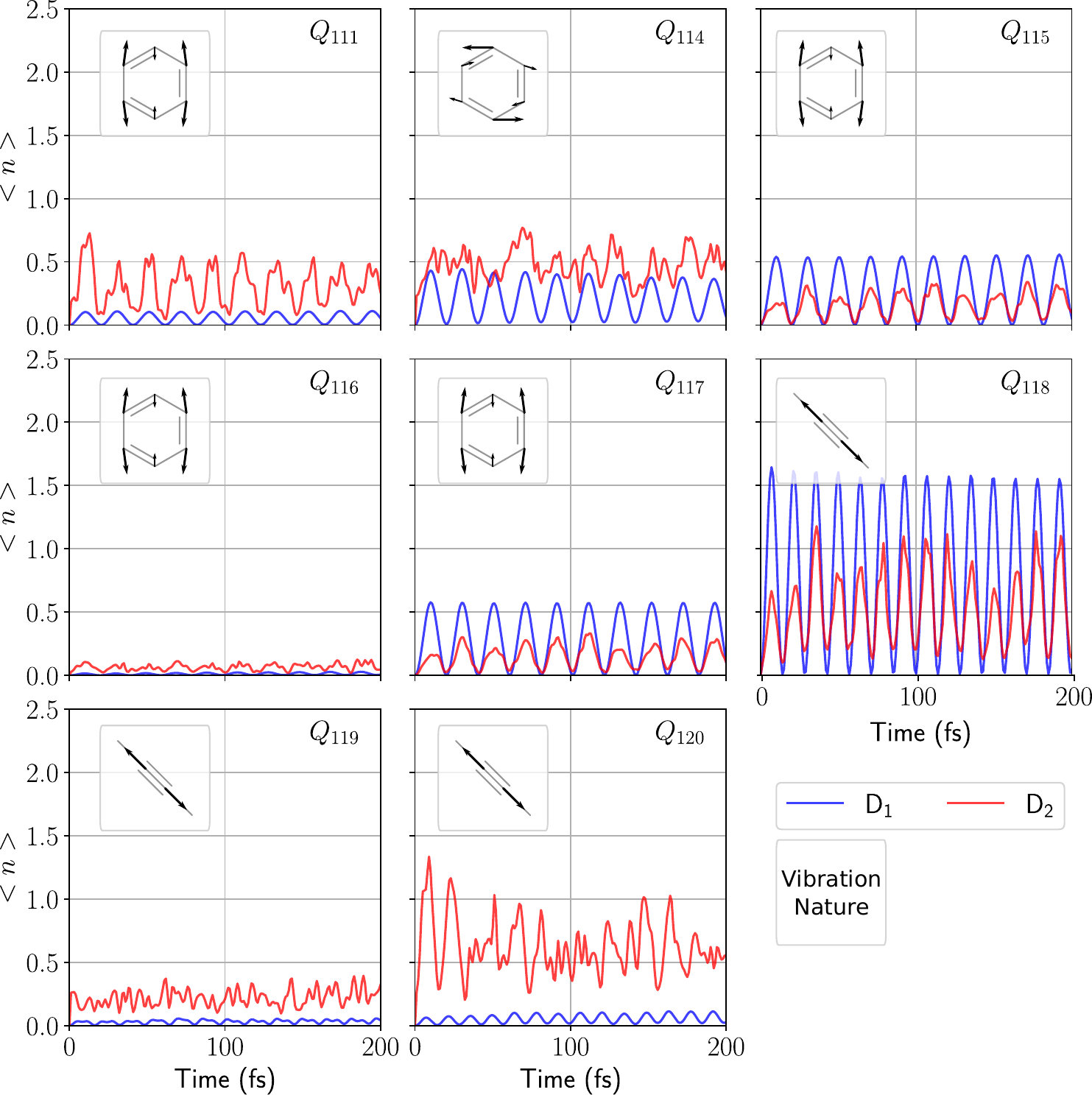}
  \caption{
    Time evolution of the expectation values of position (left, mass-weighted atomic units) and vibrational excitation number (right) in diabatic states D\textsubscript{1} (blue line) and D\textsubscript{2} (red line) for an initial excitation on D\textsubscript{1}, for each normal mode.
    The nature of the vibration (quinoidal/acetylenic and elongating/rocking) is recalled for each normal mode.
    }
  \label{fig:exc_2_QNExp_Normalized}
\end{figure}
\begin{figure}[H]
\centering
  \includegraphics[width=0.45\linewidth]{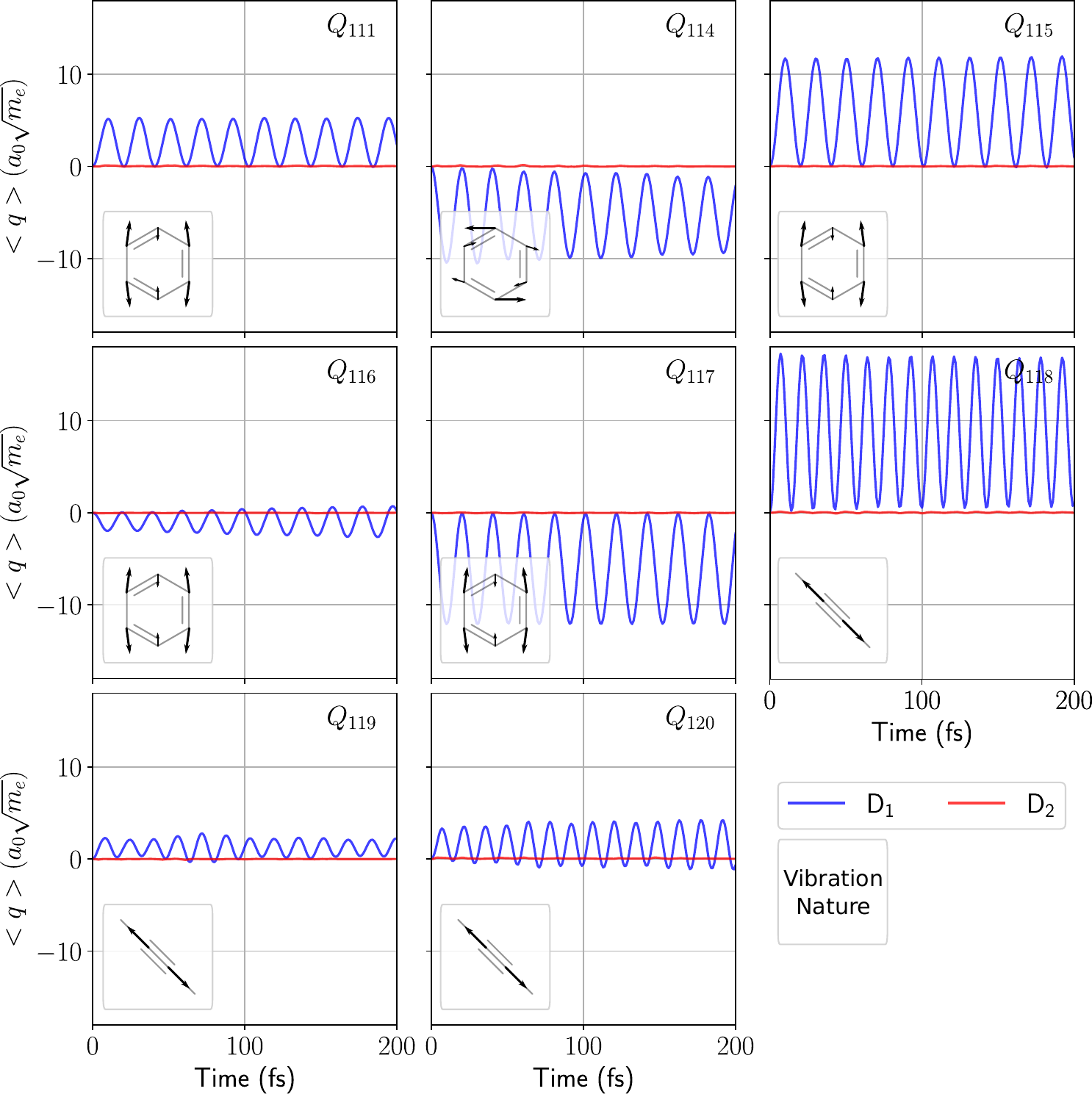}
  \includegraphics[width=0.45\linewidth]{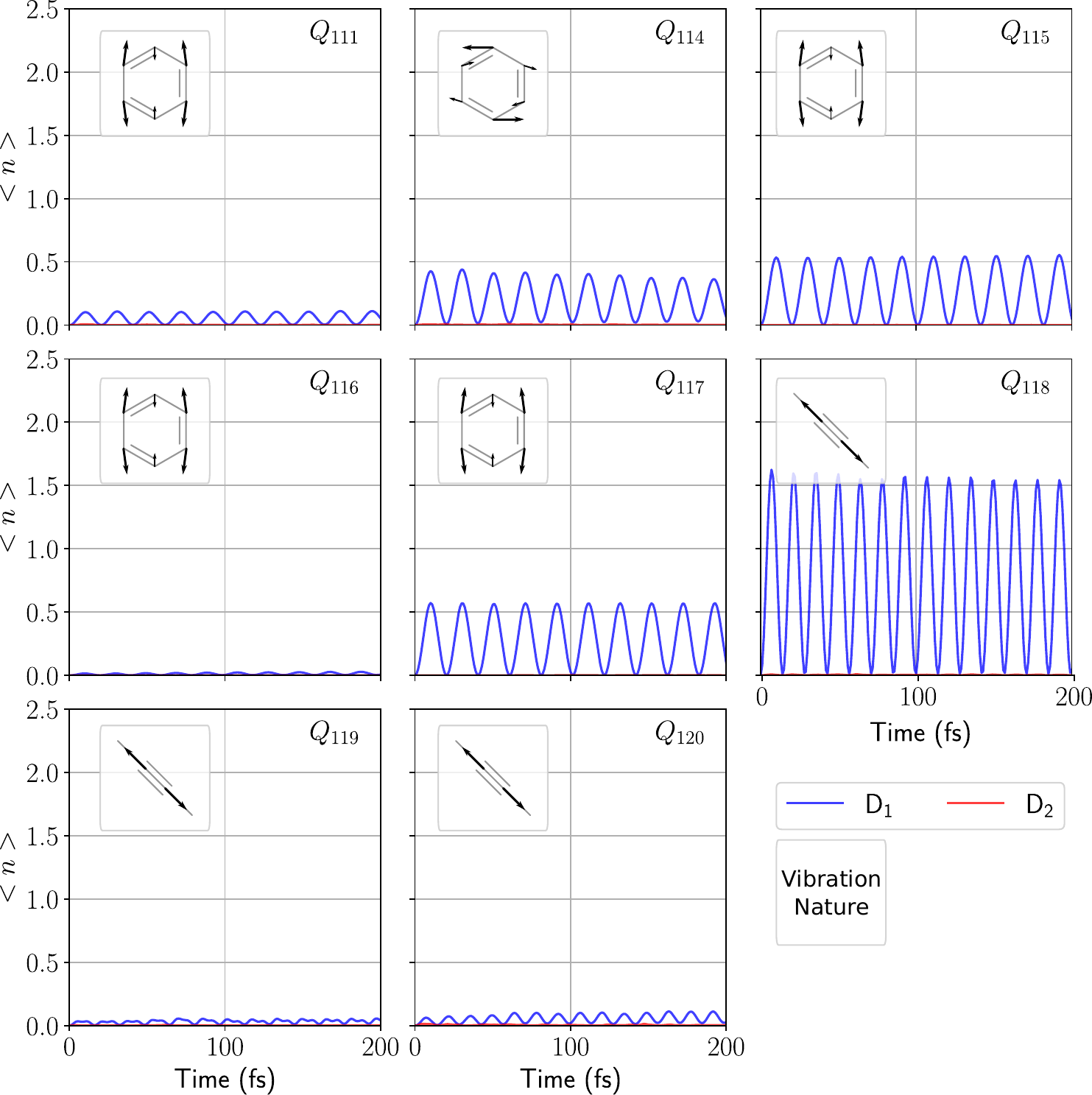}
  \caption{
    Time evolution of the population-weighted expectation values of position (left, mass-weighted atomic units) and vibrational excitation number (right) in diabatic states D\textsubscript{1} (blue line) and D\textsubscript{2} (red line) for an initial excitation on D\textsubscript{1}, for each normal mode.
    The nature of the vibration (quinoidal/acetylenic and elongating/rocking) is recalled for each normal mode.
    }
  \label{fig:exc_2_QNExp}
\end{figure}
\begin{figure}[H]
\centering
  \includegraphics[width=0.45\linewidth]{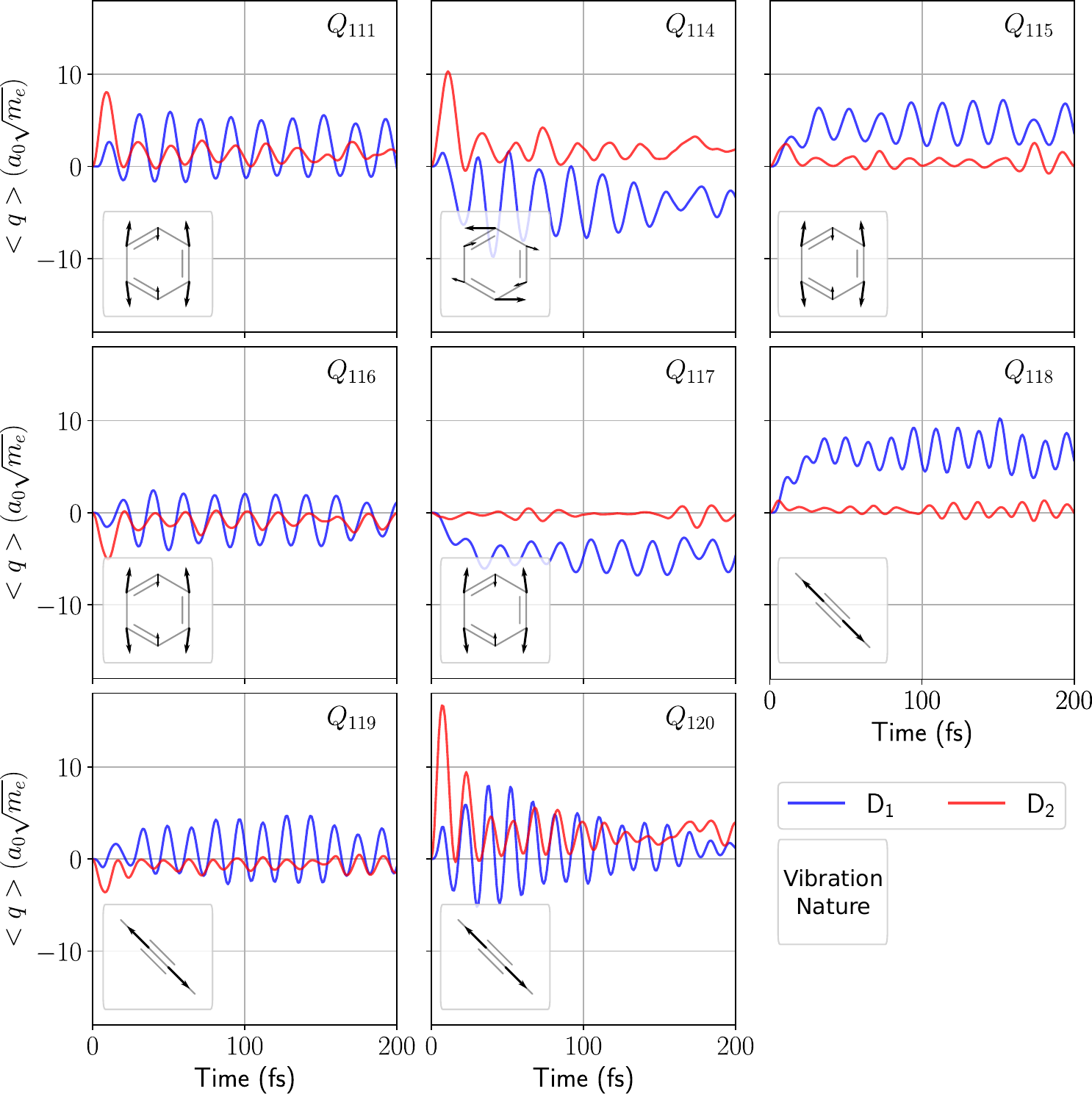}
  \includegraphics[width=0.45\linewidth]{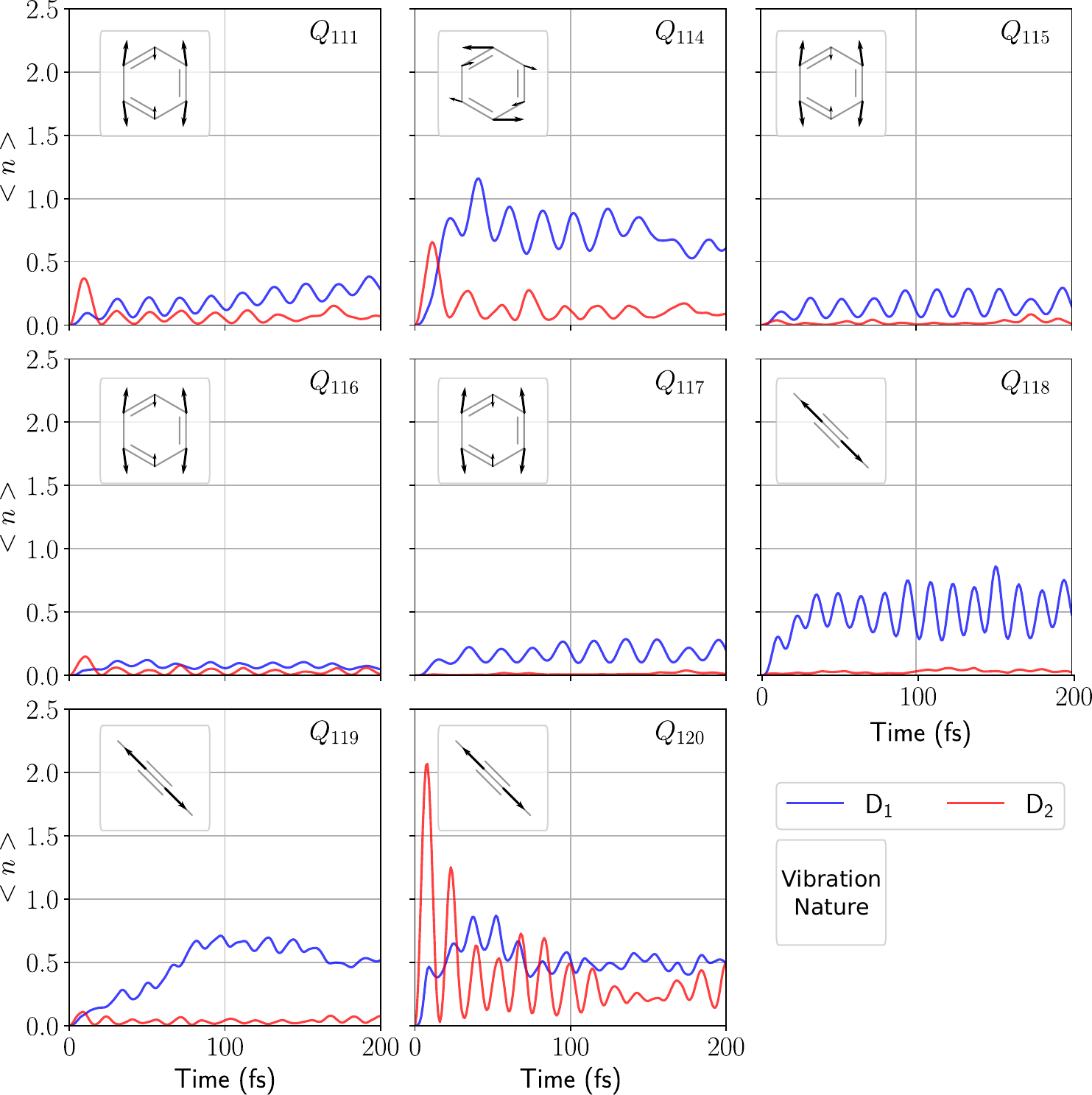}
  \caption{
    Time evolution of the population-weighted expectation values of position (left, mass-weighted atomic units) and vibrational excitation number (right) in diabatic states D\textsubscript{1} (blue line) and D\textsubscript{2} (red line) for an initial excitation on D\textsubscript{2}, for each normal mode.
    The nature of the vibration (quinoidal/acetylenic and elongating/rocking) is recalled for each normal mode.
    }
  \label{fig:exc_3_QNExp}
\end{figure}
\begin{figure}[H]
    \centering
    \includegraphics[width=0.45\linewidth]{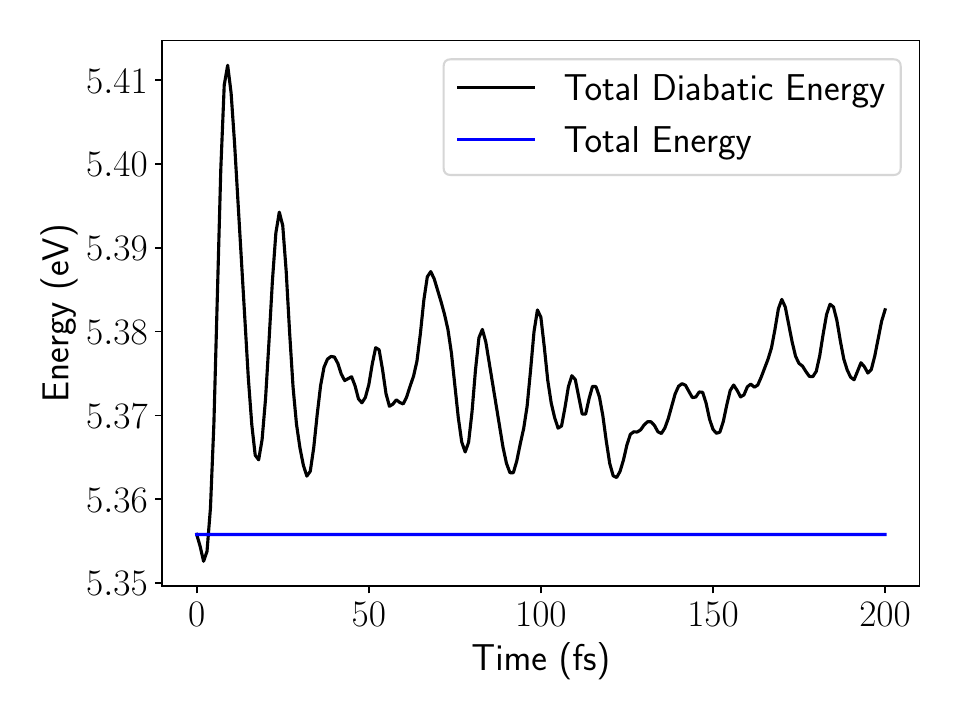}
    \includegraphics[width=0.45\linewidth]{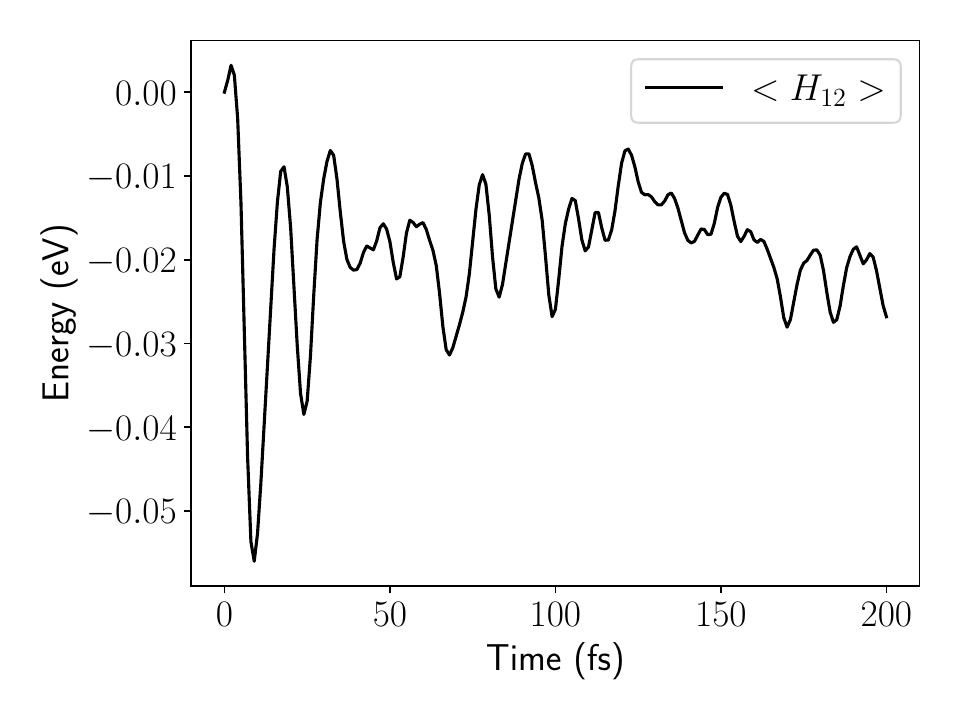}
    \caption{Total diabatic potential (left, black line) and coupling energy contribution (right) to the conserved total energy (left, blue line)}
    \label{fig:total_energy}
\end{figure}
\begin{figure}[H]
  \includegraphics[width=0.8\linewidth]{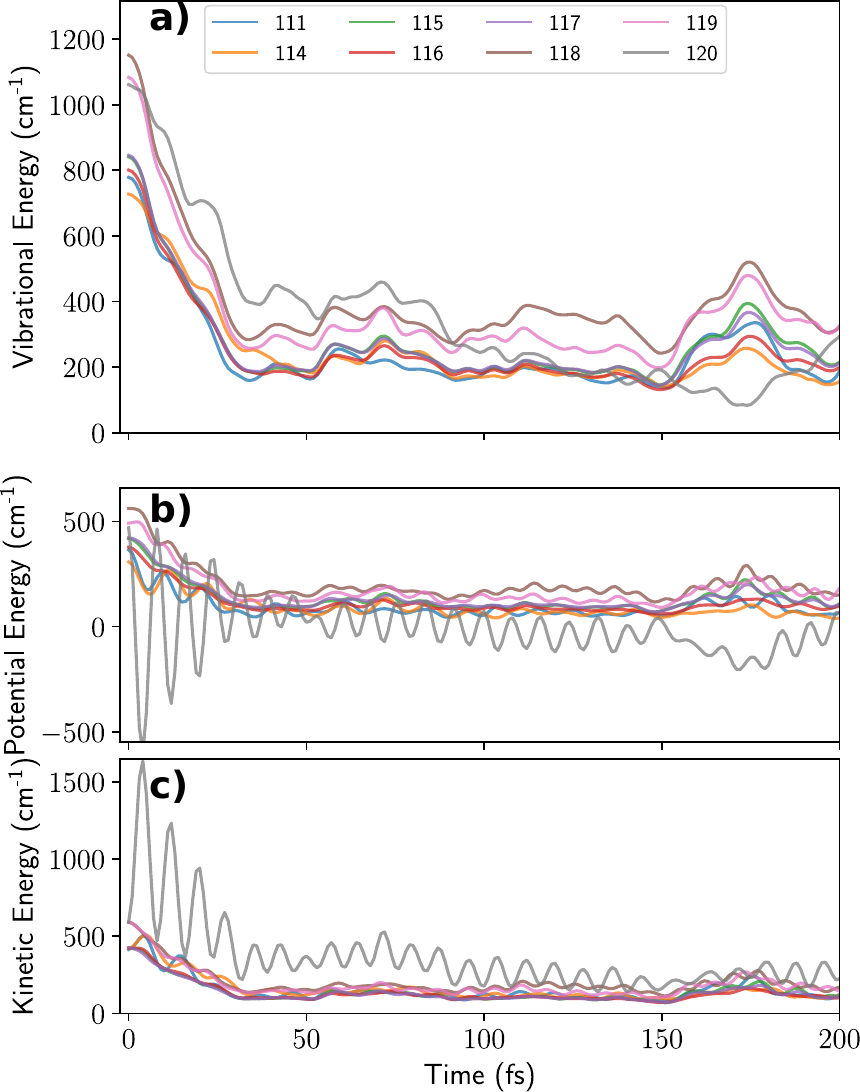}
  \caption{
    Time evolution of the vibrational, potential and kinetic energy mode per mode for contributions from diabatic states D\textsubscript{2}.
  }
  \label{fig:energy_decomposition_VibPerMode}
\end{figure}

\section{Optimized Molecular Geometries}
\subsection{Franck-Condon point, MinS\textsubscript{0} geometry}
\newpage
\noindent
\includegraphics[page=1]{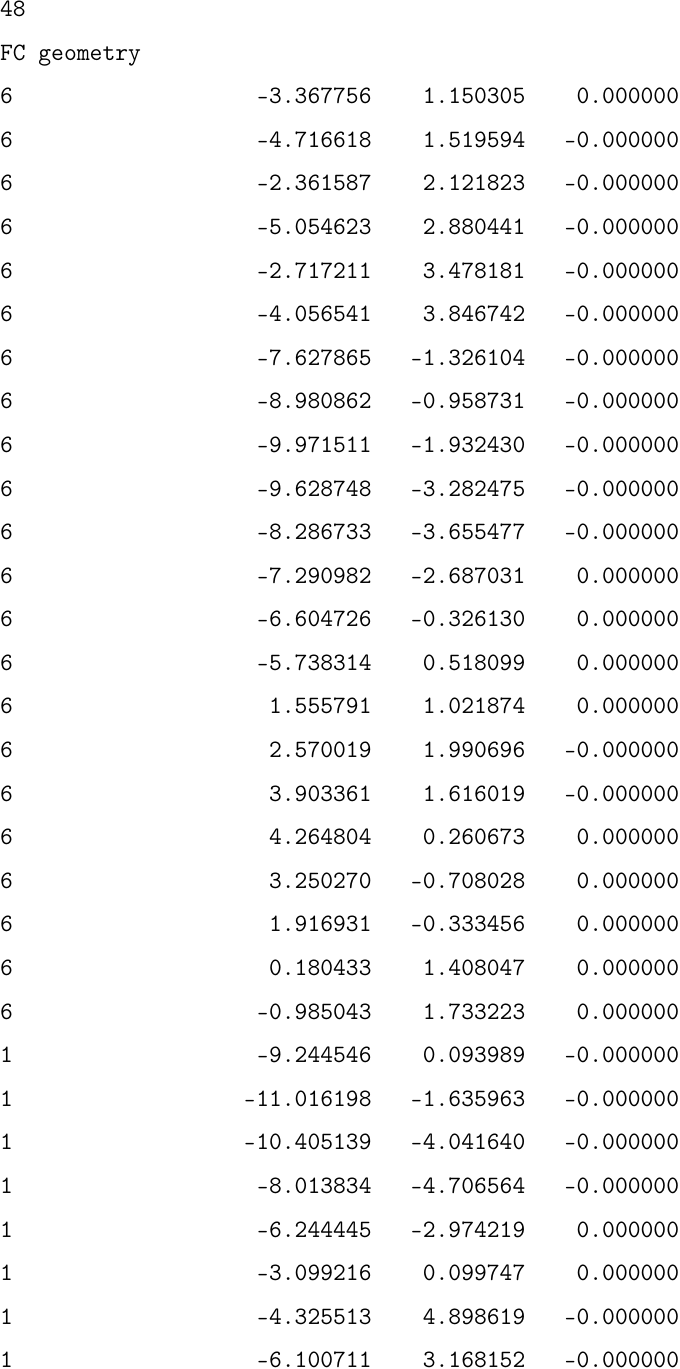}
\newpage
\noindent
\includegraphics[page=2]{FCxyz2pdf-crop.pdf}
\subsection{Minimum Energy Conical Intersection (MECI) geometry}
\newpage
\noindent
\includegraphics[page=1]{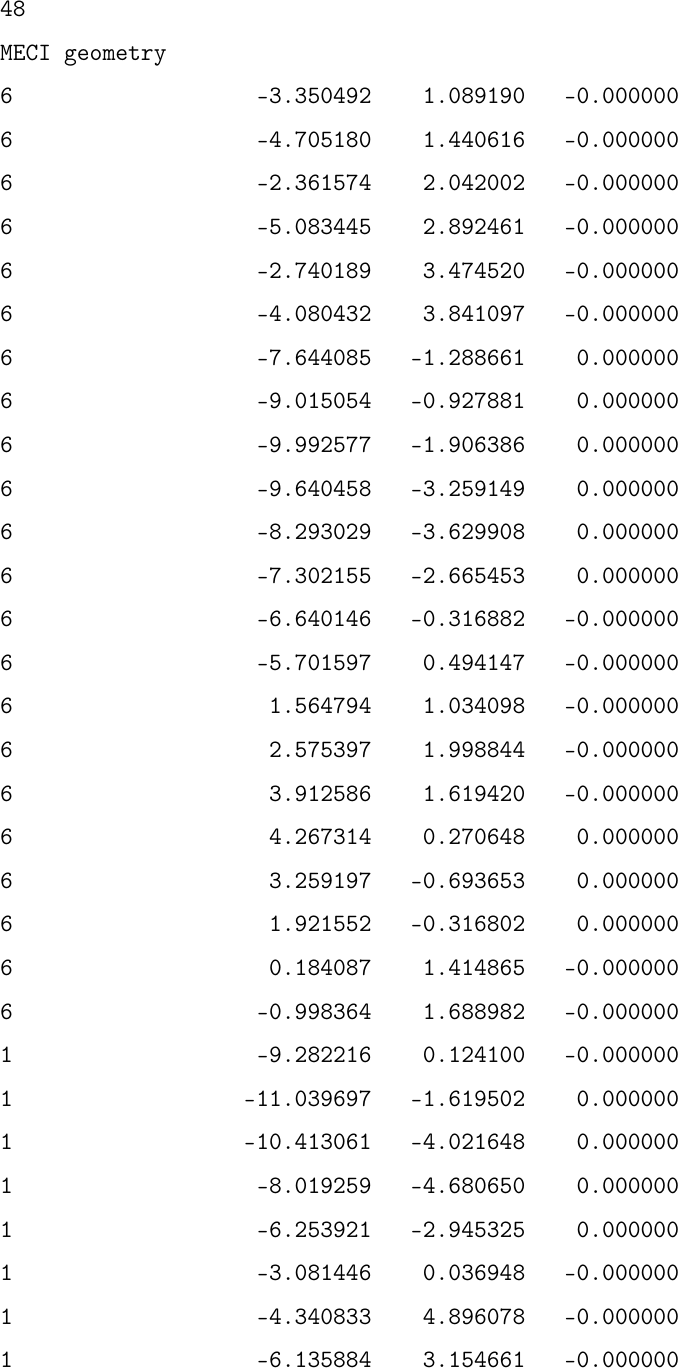}
\newpage
\noindent
\includegraphics[page=2]{MECIxyz2pdf-crop.pdf}
\section{Quantics Operator file for the LVC Hamiltonian model} 
\noindent
\includegraphics[page=1]{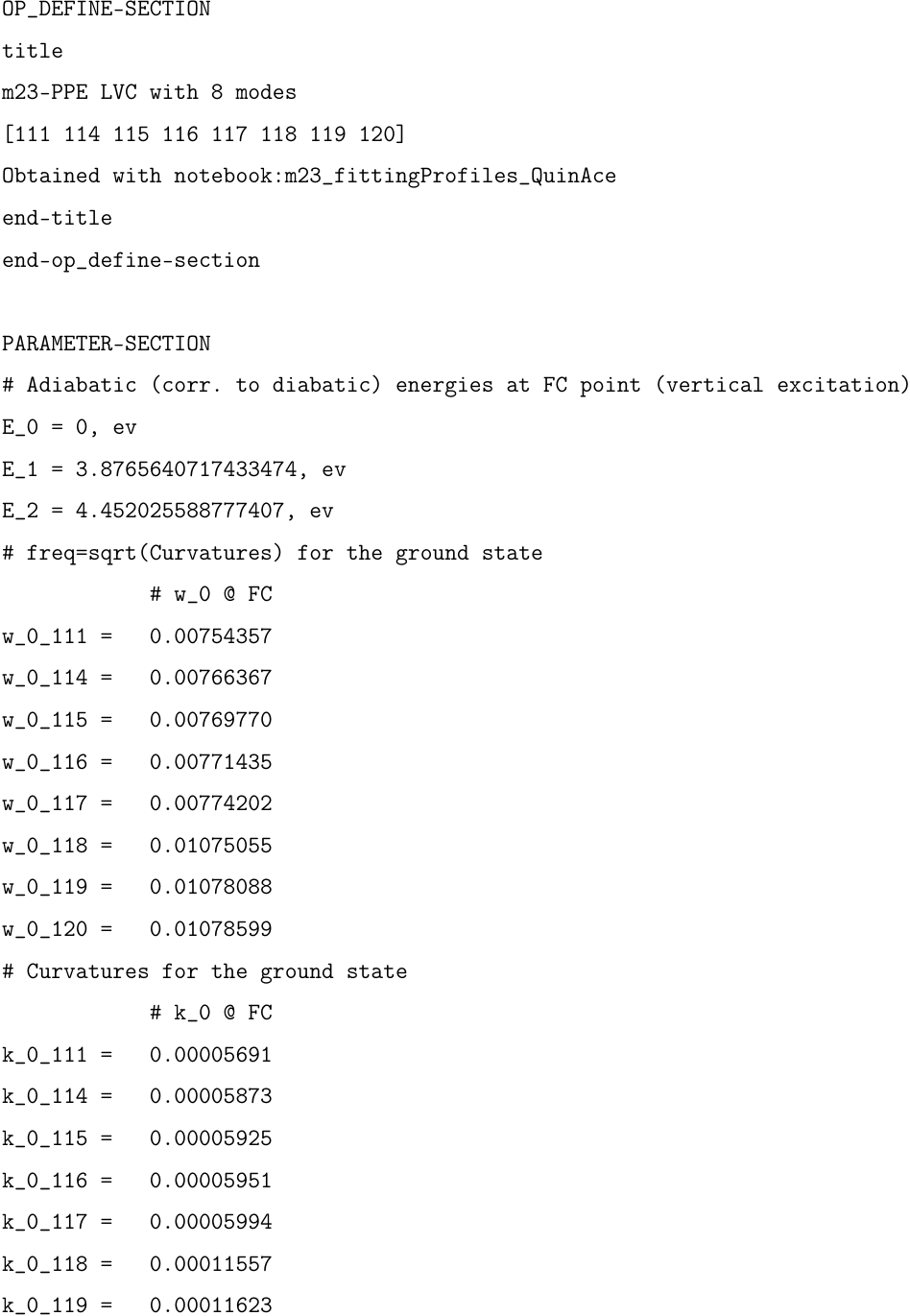}
\newpage
\noindent
\includegraphics[page=2]{op2pdf-crop.pdf}
\newpage
\noindent
\includegraphics[page=3]{op2pdf-crop.pdf}
\newpage
\noindent
\includegraphics[page=4]{op2pdf-crop.pdf}
\newpage
\noindent
\includegraphics[page=5]{op2pdf-crop.pdf}
\newpage
\noindent
\includegraphics[page=6]{op2pdf-crop.pdf}